\documentclass[preprint,11pt]{aastex}

\usepackage{natbib}
\usepackage{amssymb}		
\usepackage{subfigure}

\newcommand{\scinot}[2]{ \ensuremath{#1\times10^{#2}} }
\newcommand{\app}{$\sim$}
\newcommand{\kms}{~km\,s$^{-1}$}
\newcommand{\msol}{\,$M_{\odot}$}

\newcommand{\mrate}{\,$M_{\odot}$\,yr$^{-1}$}

\newcommand{\flux}{~erg\,s$^{-1}$\,cm$^{-2}$}
\newcommand{\surfb}{~erg\,s$^{-1}$\,cm$^{-2}$\,arcsec$^{-2}$}
\newcommand{\lum}{~erg\,s$^{-1}$}
\newcommand{\cc}{~cm$^{3}$}
	
\newcommand{\abell}{A\,3581}
\newcommand{\zw}{2A\,0335+096}
\newcommand{\sersic}{Sersic\,159-03}
\newcommand{\NIIHa}{[\ion{N}{2}]/H$\alpha$}

\newcommand{\ha}{H$\alpha$}
\newcommand{\degree}{\degr}

\shorttitle{Feedback in BCGs}
\shortauthors{Farage et al.}

\begin{document}

\title{Feedback in the cores of clusters Abell~3581, \\
2A~0335+096, and Sersic~159-03}

\author{Catherine L. Farage\altaffilmark{1}, Peter J. McGregor\altaffilmark{1}, Michael A. Dopita\altaffilmark{1}\altaffilmark{2}\altaffilmark{3}}
\email{cfarage@mso.anu.edu.au; peter@mso.anu.edu.au; mad@mso.anu.edu.au}
\altaffiltext{1}{Research School of Astronomy and Astrophysics, The Australian National University, Cotter Rd., Weston ACT 2611, Australia }
\altaffiltext{2}{Astronomy Department, King Abdulaziz University, P.O. Box 80203, Jeddah, Saudi Arabia}
\altaffiltext{3}{Institute for Astronomy, University of Hawaii, 2680 Woodlawn Drive, Honolulu, HI 96822, USA}

\begin{abstract}
The cores of massive galaxy clusters, where hot gas is cooling rapidly, appear to undergo cycles of self-regulating energy feedback, in which AGN outbursts in the central galaxies episodically provide sufficient heating to offset much of the gas cooling. 
We use deep integral-field spectroscopy to study the optical line emission from the extended nebulae of three nearby brightest cluster galaxies and investigate how they are related to the processes of heating and cooling in the cluster cores. 
Two of these systems, \abell\, and \sersic, appear to be experiencing phases of feedback that are dominated by the activity and output of a central AGN. \abell\, shows evidence for significant interaction between the radio outflows and the optical nebula, in addition to accretion flows into the nucleus of the galaxy. X-ray and radio data show that \sersic\, is dominated by the feedback of energy from the central AGN, but the kinematics of the optical nebula are consistent with infall or outflow of material along its bright filaments.
The third system, \zw, is dominated by mass accretion and cooling, and so we suggest that it is in an accumulation phase of the feedback cycle. The outer nebula forms a disk-like structure, $\sim14$\,kpc in radius, that rotates about the central galaxy with a velocity amplitude of $\sim200$\kms. Overall, our data are consistent with ongoing AGN-driven feedback cycles occurring in these systems.
\end{abstract}

\keywords{galaxies: elliptical and lenticular, cD --- galaxies: clusters: individual (\abell, \zw, \sersic) --- galaxies: ISM --- techniques: imaging spectroscopy}

\section{Introduction}\label{sec:intro}

Extended emission-line regions are a characteristic feature of the brightest cluster galaxies (BCGs) in many galaxy clusters \citep[e.g.,][]{Heckman:1981p4057,Cowie:1983p12054,Crawford:1992p3755}. These nebulae often have filamentary structures and may reach to tens of kiloparsecs from the galaxy center. Their spectra feature strong low-ionization emission lines, similar to those from the Low-Ionization Nuclear Emission-line Regions (LINERs) that are observed in many galaxy nuclei \citep{Heckman:1980p10009,Filippenko:2003p2634}.  

BCGs are giant elliptical galaxies, and commonly cD galaxies with extended stellar envelopes. They are the most luminous and most massive of the cluster galaxies and are located at, or close to, the gravitational center of the host cluster. This position contributes to a rich history of accretion and mergers, and also means they are permeated by the densest, X-ray-emitting gas of the ICM and may be the destination of any material cooling from this medium. 

In cooling-flow, or cool-core, clusters the X-ray surface brightness profile has a strong central peak, indicating a core of high density and reduced temperature. The cooling time in this core is short relative to the lifetime of the cluster, so there is the potential for massive cooling flows to have formed \citep{Fabian:1977p8925,Cowie:1977p10039,Peterson:2003p2316}. However, X-ray spectroscopy has shown that most of the gas cooling is being offset by some source of heating in the core. This source is not certainly associated with the activity of the central galaxy, but episodic outbursts of radiative and mechanical energy from active galactic nuclei (AGN) in the BCGs are currently the favored heating mechanism \citep[e.g., see][]{Begelman:2004p4286,McNamara:2007p345}. 
 
The line-emitting filaments that surround many BCGs may be symptoms of the interactions between the cooling cluster gas and the sources of mass and energy feedback in the central galaxy \citep[e.g.,][]{McNamara:2007p345}. As such, cluster cores are effective places to study feedback in nearby galaxies, through the excitation, kinematics, and origins of the nebulae. 

Observations reveal clear links between the extended emission-line regions and their environment. BCGs are more likely to host extended emission-line nebulae when they are situated close to the X-ray center of a strongly cooling cluster core \citep{Edwards:2007p2760,Cavagnolo:2008p8811}. 
In addition, BCGs that host more luminous emission-line systems tend to be found in more luminous X-ray cluster cores with higher central cooling rates \citep{Edge:2001p4022,McDonald:2010p12030,McDonald:2011p12317}, are generally associated with larger masses of cold gas \citep{Edge:2002p2508}, and have higher rates of star formation \citep{Crawford:1999p1097,ODea:2008p8568}.  
This suggests that the presence, or excitation, of the extended nebulae in cooling-flow BCGs is directly related to the cooling gas in the cluster core. 

Observational results also highlight a connection between AGN output, star formation, and gas cooling in BCG environments. For example, there appears to be a minimum threshold in central X-ray cooling rate that is required for the onset of both nuclear activity and star formation in BCGs \citep{Cavagnolo:2008p8811,Rafferty:2008p6522}. 

Models of episodic AGN heating and feedback provide at least a qualitative explanation for how the heating and regulation of cooling flows may occur \citep[e.g.,][]{Binney:1995p12530,Ciotti:2001p12528,Jones:1997p12525,Tucker:1997p12527,Pizzolato:2005p6412}. 
In this paradigm the BCG and its environment experiences cyclic episodes of cooling and inflow that feed nuclear activity, and heating and outflow powered by the AGN.
In the cold feedback model described by \citet{Pizzolato:2005p6412}, feedback occurs throughout the cool inner region of a cluster core (over radii of $\sim5-30$\,kpc). An outburst from the AGN in the central galaxy ejects a large amount of energy into the surrounding medium. Bipolar jets form and drive expanding plumes of hot, radio-emitting plasma that rise buoyantly in the cluster atmosphere. These interact with the surrounding galaxy and environment, and displace and heat the cooling cluster gas \citep[e.g., see][for a review of AGN heating mechanisms]{McNamara:2007p345}. This heating slows fuel supply to the AGN and as outflow activity decreases, material in the cluster core begins to cool. Clumps that become sufficiently dense fall inwards and, with some initial angular momentum, may form an accretion disk. This produces a reservoir of cold gas in the galaxy; some of the cooling material contributes to star formation, and some is accreted into the black hole. A large mass of accreting material will trigger another burst of nuclear activity and continue the feedback cycle.

\citet{Wilman:2009p7537} suggest that the diversity of properties of the multi-wavelength emission from cluster cores might be explained, at least in part, because they are observed at different phases of such  feedback cycles. They discuss several BCG systems and propose that one is in an early cooling phase: having gas motions in the filaments consistent with free-fall into the galaxy, weak radio emission and a strong X-ray cooling rate. They place two others, hosting stronger and more extensive radio sources,  in post-AGN outburst phases.  
Such studies of the multi-phase gas surrounding BCGs, particularly including integral-field spectroscopic data that provide kinematic information about the emission-line regions, will help us to understand which, if any, properties of the extended emission systems are evidence of feedback activity. Further, this may allow us to definitively place individual BCG systems within the sequence of a heating/cooling cycle, and potentially to constrain the physics of feedback models in clusters. 

We have previously presented a deep, integral-field spectroscopic study of NGC~4696, the central galaxy of the Centaurus cluster, that provided a detailed picture of the kinematics of the central emission-line filaments \citep{Farage:2010p11714}. This system may be in a period of accretion, following the infall of a small cluster galaxy. We found that models of \app200\kms\, shocks, propagating through pre-ionized, dense clouds, can reproduce the optical emission-line spectrum. The results also emphasize that mergers and galaxy interactions are an important component of the activity in cooling-flow BCG nebulae.

Here, we study a small sample of other low- and intermediate-luminosity cooling-flow BCG filament systems. These are the first optical integral-field spectroscopy of the nebulae surrounding the BCG of \abell\, and \sersic, and the first to cover the full extent of the emission-line regions in that of \zw.
We investigate the kinematic and excitation properties of the optical emission surrounding the galaxies and combine our results with those of studies at X-ray and radio frequencies to make inferences about the state of feedback activity in these systems.

Throughout this study we adopt the cosmological parameters $H_{0} = 71$\,km\,s$^{-1}$\,Mpc$^{-1}$, $\Omega_{M} = 0.27$ and $\Omega_{\Lambda} = 0.73$, based on the five-year \textit{Wilkinson Microwave Anisotropy Probe} (\textit{WMAP}) results \citep{Hinshaw:2009p10650}. 

\section{Brightest cluster galaxy sample}\label{sec:sample}

We have observed three brightest cluster galaxies that are known to host extended nebulae that produce strong low-ionization optical emission lines; e.g., [\ion{N}{2}], [\ion{O}{2}], [\ion{S}{2}], [\ion{O}{1}], and [\ion{O}{3}] \citep{Johnstone:1987p4824, Romanishin:1988p11445,Crawford:1999p1097}. They are all well-known examples of such systems.

The galaxies represent the low-to-intermediate end of the range of total emission-line luminosity of BCGs. 
The {\em ROSAT} Brightest Cluster Sample \citep[an X-ray-selected sample of BCGs from the {\em ROSAT} All-Sky Survey;][]{Allen:1992p1098} covers a range of integrated H$\alpha$ luminosities from $10^{39}$ to $10^{43}$\lum\, \citep{Crawford:1999p1097}. The BCGs in \abell, \zw, and \sersic\, have H$\alpha$ luminosities of \scinot{\sim(5-50)}{40}\lum. 
Table~\ref{tab:sample} compares some observed properties of these galaxies, together with NGC~4696 \citep{Farage:2010p11714}. Throughout the rest of this paper we refer to the central galaxies by the host cluster names listed in this table.  


\subsection{Abell 3581}
\abell\, is a relatively nearby ($z\sim0.022$) cluster, however the extended nebula that surrounds the BCG has been little studied since early imaging and spectroscopy by \citet{Robinson:1987p317} and \citet{Danziger:1988p8014}.
The radio and X-ray emission in the cluster core have been investigated in some detail, though, using observations from the \textit{VLA}, \textit{XMM-Newton}, and \textit{Chandra} \citep[e.g.,][]{Johnstone:1998p7726,Dunn:2004p6461,Johnstone:2005p7725,Sanders:2010p11151}. 

The BCG is a strong radio source, identified as PKS~1404-267. Two irregular lobes of 1.4\,GHz emission extend \app10\arcsec\,(4.6\,kpc) to the east and west of a compact central source that coincides with the optical centroid of the galaxy \citep{Johnstone:1998p7726}. 
Both radio lobes appear to have suffered some disruption: there are distortions or bends in the orientation of the extended structures. 

The 1.4\,GHz radio lobes coincide with two cavities in the hot X-ray halo gas of the cluster core \citep{Johnstone:2005p7725}. Such structures are seen in many BCG cooling-flow systems and are assumed to be evidence that the radio lobes displace the surrounding hot gas as they expand and propagate away from the central galaxy \citep[see][]{McNamara:2007p345}.  
\citet{Rafferty:2006p2278} calculated the total power required to inflate the cavities seen in the X-ray-emitting ICM of \abell. 
They found that the cavity power and cooling luminosities approximately balance in this system. The ratio of cavity power to the cooling X-ray luminosity in the core is: $P_{\rm cav} / L_{\rm ICM} \sim 0.8^{+1.0}_{-0.4}$. 
X-ray spectroscopy indicates a relatively low rate of gas cooling to low  temperatures in the core: $\dot{M}<10$\msol\,yr$^{-1}$ \citep{Voigt:2004p6445,Sanders:2010p11151}.
The X-ray and radio observations suggest that most of the ICM cooling is offset by the current AGN feedback activity in this system.

\subsection{2A 0335+096}
\zw\footnote{~identified as RX~J0338.7+0958 in some studies} is a compact cluster with a luminous X-ray core \citep{Sanders:2009p11794}. The BCG has cD morphology, hosts a weak radio source, and is surrounded by an extensive optical emission-line nebula that was first imaged by \citet{Romanishin:1988p11445}. They reported a clumpy, 12\,kpc-long bar of line emission and a 2\,kpc disk-like feature that they characterized as possible accretion disk inclined less than 17\degr\, from edge on.

In optical integral-field spectroscopy of the central 7\arcsec\,$\times\,$9\arcsec\, of the nebula, \citet{Hatch:2007p2057} resolved two knots of emission within the inner bar of \ha\, emission. Both knots are less than $\sim2$\arcsec\, (1.4\,kpc) from the galaxy core and show distinct line-of-sight velocities separated by $\sim200$\kms. These structures were also seen in detailed near-infrared integral-field observations of the central 3\arcsec\,$\times\,$3\arcsec\, by \citet{Wilman:2011p13215}. 
A more diffuse region of emission extends from the northwest knot towards a nearby galaxy, $\sim6$\arcsec\, (4\,kpc) away on the sky to the northeast of the nucleus. The line-of-sight velocity increases smoothly from $\sim-250$\kms\, approximately 1\arcsec\, south of the BCG nucleus to $\sim200$\kms\, at the position of the companion galaxy. \citet{Hatch:2007p2057} suggest that the secondary galaxy may have disturbed a reservoir of molecular gas in the BCG to produce this extension of the inner nebula towards the northwest. 

Long-slit spectroscopy by \citet{Gelderman:1996p11337} and \citet{Donahue:2007p11428} provided evidence that the large-scale bar of line emission that extends over more than 6\,kpc to the northwest and southeast is kinematically distinct from the inner emission structure.  The line-of-sight velocity difference between the outer regions of the large-scale bar is reversed with respect to the velocity gradient across the inner disk between the galaxies. \citet{Donahue:2007p11428} noted that a two-dimensional velocity map is needed to characterise the velocity field of the nebula.

In 1.4\,GHz and 5\,GHz radio images of the system, two diffuse, amorphous lobes of radio emission extend $\sim10-15$\arcsec\, ($\sim7-10$\,kpc) from galaxy center along a position angle of $60\degr$ \citep{Sarazin:1995p6240,Donahue:2007p11428,Sanders:2009p11794}. \citet{Sarazin:1995p6240} suggest that these lobes are associated with oppositely-directed radio jets and have been disrupted by the high-pressure ambient gas.  \citet{Sarazin:1995p6240} measure a spectral index of $\alpha\sim-0.3$ on the galaxy nucleus, and $\alpha\sim-1$ in the two radio lobes.  Based on steepening as a result of synchrotron losses (from $\alpha\sim-0.5$), and assuming an equipartition magnetic field, \citet{Donahue:2007p11428} estimated an age of $25-50$\,Myr for the radio lobes. They proposed that an interaction with the nearby companion elliptical galaxy may have triggered the current episode of radio activity. 

There is evidence for an active history of interaction between the AGN and ICM in the core of \zw. Studies with \textit{XMM-Newton} and \textit{Chandra} have identified a complex system of X-ray-emitting structures in the core, including filaments, cool clumps, a metal-rich spiral, and at least five distinct cavities at varying distances and position angles relative to the BCG \citep{Mazzotta:2003p11448,Birzan:2004p8008,Kaastra:2004p10623,Sanders:2006p6787,Sanders:2009p11794}.  The radio lobes are associated with weak cavities in the X-ray gas, but the most prominent cavities coincide with very extended, diffuse lobes of 1.4\,GHz emission approximately $20-30$\,kpc from the galaxy core on the sky. The complex system of cavities presumably originates from multiple previous generations of AGN radio outbursts \citep{Sanders:2009p11794}. 

X-ray observations indicate a relatively high mass cooling rate in the core: $30-80$\msol\,yr$^{-1}$ \citep{Rafferty:2006p2278,Voigt:2006p11452}, and a high ratio of X-ray cooling luminosity to radio power output \citep{Rafferty:2006p2278,Birzan:2004p8008}.  The morphology of the coolest X-ray-emitting material is correlated closely with the large-scale bar of optical line emission \citep{Sanders:2009p11794}, so there is cooling X-ray gas associated with extended nebula. This system currently appears to be dominated by cooling rather than AGN activity.

\subsection{Sersic 159-03}
\sersic\footnote{~also designated Abell S1101 (A S1101)} is a relatively low-mass cool-core cluster that has been the subject of many observational studies. The core of the cluster has a complex structure at all wavelengths and shows signs of strong AGN feedback \citep{Werner:2011p12464,Birzan:2008p6520}. 

Previous imaging \citep{Hansen:2000p9614,Jaffe:2005p2500,Crawford:1992p3755,McDonald:2010p12030,Oonk:2010p11713,McDonald:2010p12030,McDonald:2011p12317,Werner:2011p12464} in the optical and near-infrared shows several bright filaments surrounding the BCG. A curved filament extends north of the BCG nucleus to a radius of $\sim35$\,kpc, and comprises a number of filaments and clouds. 
\textit{Hubble Space Telescope} (\textit{HST}) observations of the galaxy revealed two lanes of dust absorption that are associated with the emission-line structures \citep{Werner:2011p12464}. Faint ultraviolet emission is also extended along the filament. The presence of ultraviolet radiation and dust may be evidence for star formation associated with this structure.

\sersic\, is associated with a complex radio source.  Two inner lobes of radio emission are seen at 5\,GHz and 8.5\,GHz \citep{Birzan:2008p6520,Werner:2011p12464}. The 5\,GHz emission is centrally concentrated and the relatively faint lobes are irregular and asymmetric. However, the radio emission is considerably more extended at lower frequencies. Emission at 300\,MHz is detected at radii of over an arcminute \citep{Birzan:2008p6520}. 
The radio source has a very steep power-law spectrum ($\alpha\sim-2.2$) in the extended emission to the east, northeast, and northwest of the core, and is more shallow in the nucleus  \citep[$\alpha\sim-1$ to $-1.5$;][]{Birzan:2004p8008,Werner:2011p12464}.  The most recent radio output (assumed to be the high-frequency emission) is extended to a radius of $\sim10$\,kpc in each lobe. \citet{Birzan:2004p8008} suggest that the radio source has either restarted, or is fading away.

\cite{Birzan:2008p6520} also identified indistinct, low-contrast X-ray cavities that lie beyond the emission-line filaments and are associated with diffuse, low-frequency radio emission. The power associated with inflating these cavities is high, approximately three to four times the luminosity of the cooling X-ray gas in the cluster core \citep{Rafferty:2008p6522}. 
X-ray spectroscopy shows that the ICM in the core is cooling to low temperatures at a rate of only $\lesssim30$\msol\,yr$^{-1}$ \citep{Werner:2011p12464,McDonald:2010p12030,Rafferty:2006p2278,Voigt:2004p6445}.

In a recent multi-wavelength study of the BCG core, \citet{Werner:2010p11085} found evidence for strong interaction between the AGN and hot gas of the cluster. They suggested that AGN feedback activity has removed most of the cooling X-ray gas from the core (within $r\lesssim7.5$\,kpc). They also proposed that the filamentary nebula has been uplifted from the central galaxy by the radio lobes, but that gas in the filaments is now largely removed from the impact of the radio jet and is cooling and forming stars.


\section{Observations and data reduction}\label{sec:obs}
The galaxies were observed with the Wide-Field Spectrograph \citep[WiFeS;][]{Dopita:2007p6557,Dopita:2010p10019} and the Australian National University (ANU) 2.3\,m telescope at Siding Spring Observatory. WiFeS is a double-beam, image-slicing, integral-field spectrograph that records optical spectra over a contiguous $25\arcsec\times38$\arcsec~field of view. This spatial field is divided into twenty-five 1\arcsec-wide slitlets, sampled by 0.5\arcsec\, pixels along the 38\arcsec\, length. 

The instrument is a dual-channel device with separate gratings and cameras for the blue and red channels. Each camera is equipped with 4k$\times$4k pixel CCD detectors. The low-resolution grating configuration, employing the R3000 and B3000 gratings and the RT560 dichroic, was used for the observations described here. This provides a continuous wavelength coverage from 3290 to 9330\,\AA\, in the reduced data. The spectral sampling is 0.81\,\AA\, per pixel in the blue data cubes and 1.3\,\AA\, per pixel in the red data cubes. The typical velocity resolution achieved is $\sim100$\kms\, FWHM.

The pattern of science observations consisted of two pairs of 30 minute exposures on a galaxy separated by a 30 minute exposure of the sky. The scaled sky frame was subtracted from each of the four galaxy frames in this sequence. A dither pattern of 2\arcsec\, offsets was applied to the position of each galaxy frame. A nearby reference star was used to autoguide the galaxy exposures. Table~\ref{tab:obs} shows the total integration times obtained for each galaxy and other details of the observations. 

Each set of four galaxy frames and one sky frame was preceded and followed by an observation of a spectrophotometric standard star and these were used to calibrate the absolute flux scale of the observations. Dwarf stars with spectra that resemble a featureless blackbody were also measured during the night and used to remove atmospheric absorption features from the flux standard star and galaxy spectra. 
Cu-Ar and Ne-Ar arc calibration lamp frames, for wavelength calibration, were observed regularly during each night, and quartz-iodine lamp flat-field frames were observed at the beginning of each night. Twilight sky flat-field frames were obtained on most nights, so that those used in processing an image were never obtained more than two nights from that observation.


The data were reduced using IRAF data-reduction scripts for WiFeS \citep{Dopita:2010p10019} that are based on and directly employ scripts from the Gemini IRAF package\footnote{\url{http://www.gemini.edu/sciops/data-and-results/processing-software}}.
The process of data reduction applied to these data is similar to that described in \citet{Farage:2010p11714}, to which we refer the reader for a more detailed description, but differs in several points that are outlined below.

A slightly different method was employed to remove the detector bias level from the images. Subtracting the bias signal is complicated both by detectors that are read out through four separate amplifiers, producing a different bias level in the four quadrants of each image, and by a residual shape that is present in the bias level across the detectors. These were removed using polynomial fits to the bias signal across unexposed rows on each amplifier in addition to the overscan columns.

A second-order correction for extinction in the Earth's atmosphere is also included in the flux calibration for these data. We estimate that the resulting absolute flux scale is accurate to approximately 11\%.

Also, for these data a correction for differential atmospheric refraction was made in software after constructing the data cubes, using routines from the Starlink positional astronomy library \textsc{SLALIB} in IRAF.

The average rms error in the wavelength calibration is $\sim0.2$\,\AA\, for the red cubes and $\sim0.1$\,\AA\, for the blue cubes.
Line-of-sight velocities are presented relative to the local standard of rest, as the corrections for the relative velocities of the sources during the observations were smaller than the spectral resolution of the data.

After forming the individual data cubes for each exposure, the pixels were binned by two in the spatial direction along the slitlets, to form square 1\arcsec\,$\times$\,1\arcsec\, pixels. This improves the signal-to-noise ratio and provides a spatial pixel scale that is better suited to the typical seeing during the observations ($\sim2\arcsec-2.5$\arcsec). 

Several exposures of \abell\, and \zw\, were obtained with the field oriented at a different position angle (PA) to the majority of the observations. The binned cubes were rotated to the intended position angle using the IRAF \textsc{rotate} task, interpolating the pixels linearly. The position angles for the combined data cubes are provided in Table~\ref{tab:obs}.

The individually-reduced science cubes were median-combined after applying spatial offsets to align the dithered observations. These offsets were derived from the continuum centroid of the galaxy nucleus, measured in an image close to the center of the spectrum in a region that is free of line emission. Cosmic rays and bad pixels were also removed by combining the data cubes in this way.
 
\subsection{Continuum subtraction}

\begin{figure}
\epsscale{1.}\begin{center}
\plotone{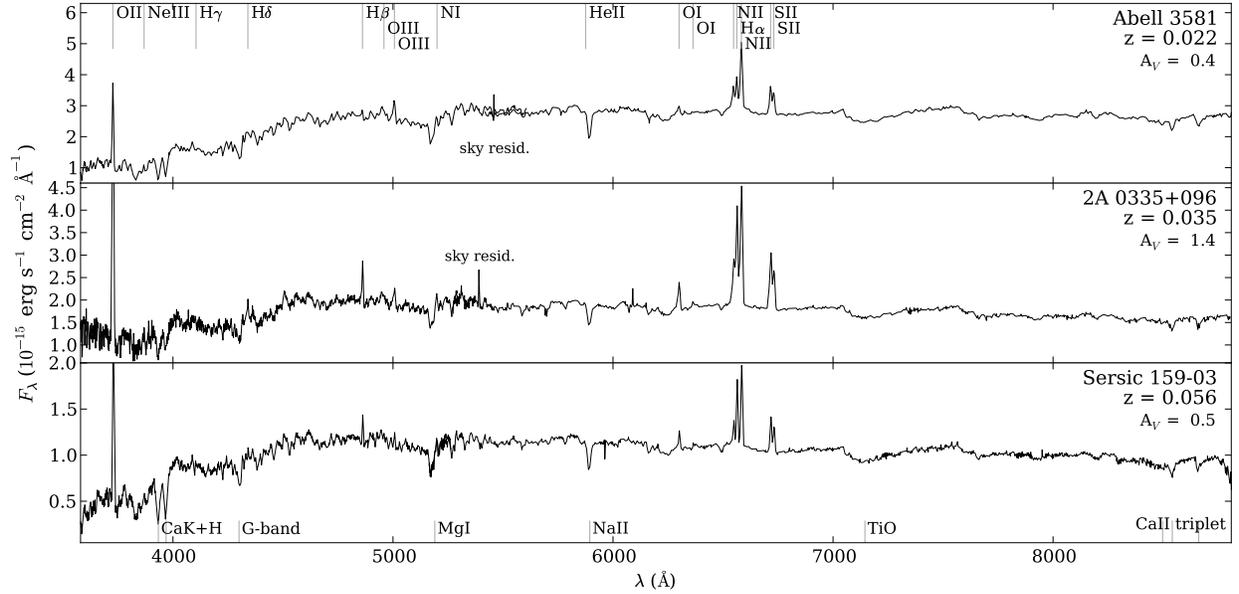}
\caption{Integrated central spectra of \abell\, (top), \zw\, (center), and \sersic\, (bottom), summed over 5\arcsec-diameter apertures centered at the peak of the BCG continuum emission in each data cube. The spectra are corrected for the level of extinction indicated in each panel, derived as described in Section~\ref{sec:linefit}.}\label{fig:intspec}
\end{center}
\end{figure}

As seen in Figure~\ref{fig:intspec}, there is a strong contribution to the BCG spectra from the stellar component of the elliptical galaxies. In the central regions of the galaxies, the spectrum of each is dominated by this emission, which must be removed to accurately measure the nebular emission lines. 


For \abell, this was done using a spectrum extracted from within the data cube itself, since line emission is not observed in all regions spatially across the galaxy. We extracted an integrated continuum template spectrum from a position 8\arcsec\, north and 5\arcsec\, west of the continuum peak, within a 4\arcsec-diameter aperture. 

We could not extract a template galaxy continuum spectrum from within the \sersic\, and \zw\, data cubes that was free of line emission and had sufficient continuum signal to be useful in removing the galaxy continuum. In these cases, we formed a galaxy continuum template spectrum from previous WiFeS observations of the giant elliptical galaxy NGC~4696. An integrated spectrum was extracted from a region of high galaxy signal and the line emission was removed using the Gaussian line-profile fits to the emission in the continuum-subtracted NGC~4696 cubes as described in \citet{Farage:2010p11714}. This spectrum was then interpolated onto a redshifted wavelength scale for each of the galaxies to form a continuum template spectrum.

\begin{figure}
\epsscale{0.95}
\plotone{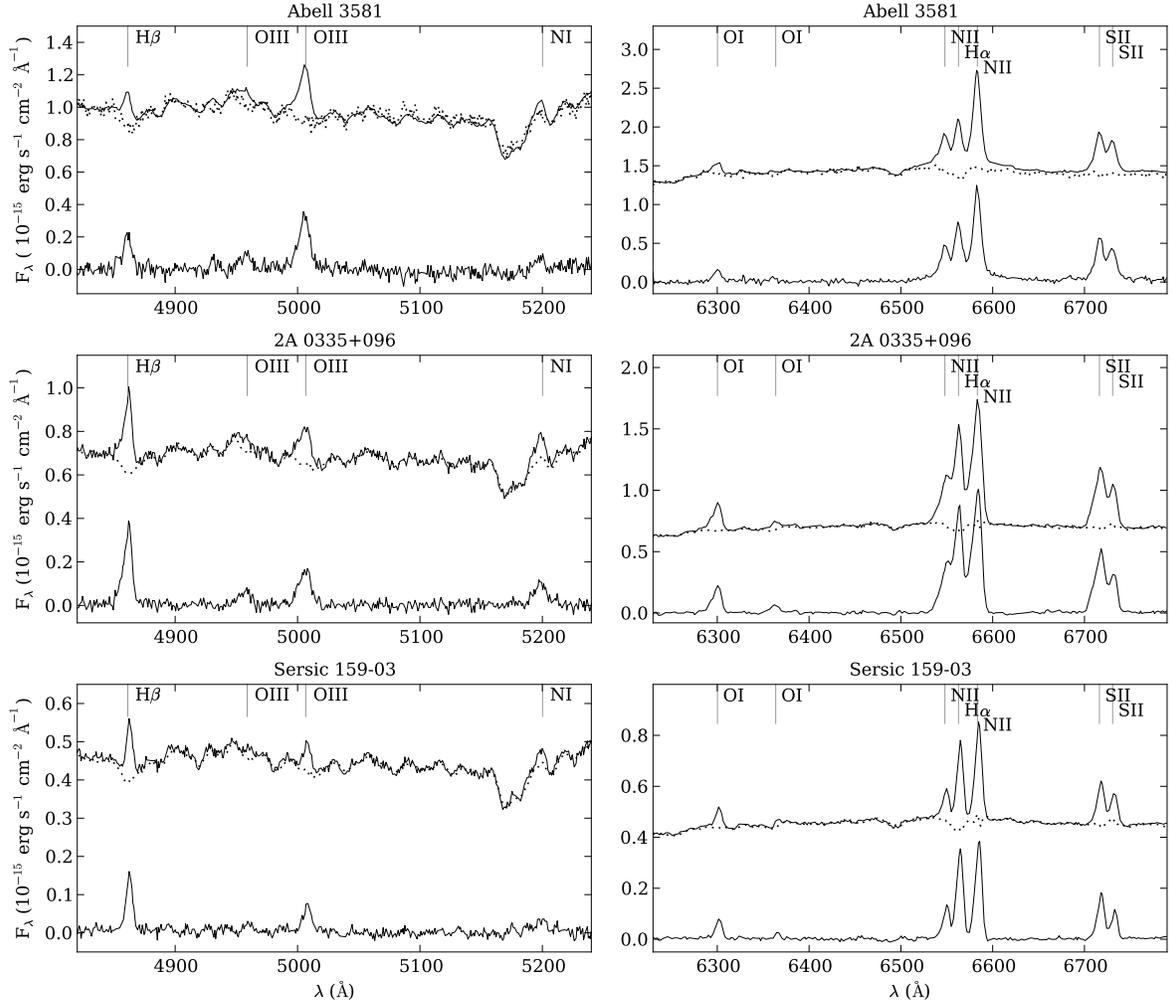}
\caption{Integrated and continuum-subtracted spectra from the cores of the BCGs; from top to bottom: \abell, \zw, and \sersic, showing spectral regions around the H$\beta$ and [\ion{O}{3}] lines in the blue data cube (left) and H$\alpha$ and [\ion{N}{2}] lines in the red cube (right). In each panel the upper solid line is the integrated galaxy spectrum, the dotted line is the continuum template spectrum, scaled to match the continuum within the displayed region of the spectrum, and the lower line is the continuum-subtracted result. Each spectrum is extracted from a 4\arcsec-diameter aperture at the galaxy nucleus.}\label{fig:contsubspec}
\end{figure}

For each of the emission lines of interest, in each spatial pixel the template spectrum was scaled to match the flux level of a continuum region close to the line and then subtracted.
Figure~\ref{fig:contsubspec} shows example continuum-subtracted spectra from the red and blue data cubes for each galaxy. The spectra are extracted from a $4$\arcsec-diameter aperture located at the galaxy nucleus, as defined by the peak of the continuum emission.

Employing continuum template spectra from the WiFeS data was the most straightforward option for removing the stellar continuum, and provides satisfactory results since the continuum shape is similar in each of these galaxies.


\subsection{Emission line profile fitting}\label{sec:linefit}

The emission-line features in the continuum-subtracted cubes were fit using the Python NMPFIT package, an implementation of Craig Markwardt's \textsc{MPFIT} routines for IDL \citep{Markwardt:2009p11918}, which applies the Levenberg-Marquardt method for least-squares minimization of data to a model. 
The line-profile models consist of one or more Gaussian functions plus a linear continuum baseline to remove any residual continuum level and slope. We constructed and fit the models separately for each of the individual emission lines or set of closely-spaced lines, since the continuum subtraction is applied  locally for each feature or set of features within the cubes.  

The \ha\, and [\ion{N}{2}]\,$\lambda\lambda$6548, 6583 doublet were fit together. We constrained the velocity of \ha\, and [\ion{N}{2}]\,$\lambda$6548 to be the same as that of [\ion{N}{2}]\,$\lambda$6583, and the ratio of the [\ion{N}{2}] doublet line amplitudes to be 3.0. These lines were also constrained to have equal velocity widths. 
The components of the [\ion{O}{3}]\,$\lambda\lambda$4959, 5007 doublet were fit together and the 5007/4959\,\AA\, amplitude ratio fixed at a value of 2.88. Both lines were also constrained to have the same relative velocity centroid and line width. 
The model of the [\ion{S}{2}]\,$\lambda\lambda$6716, 6731 doublet also consisted of two Gaussian components that have the same velocities and line widths. 
The components of the [\ion{O}{2}]\,$\lambda\lambda$3726, 3729 doublet are sufficiently broad that they are unresolved in the data cubes and a single Gaussian component was fit to this line profile everywhere. The line width is broadened by including the two components of the doublet in the single fit. A single Gaussian component was also fit to the unresolved [\ion{N}{1}]\,$\lambda\lambda$5198, 5200 doublet.

The reported line widths have been corrected for instrumental broadening of 100\kms\, FWHM, which is the mean line width measured from arc lamp frames observed with these data. Line widths are given as velocity dispersions, rather than FWHM (full width at half maximum). 
Line-of-sight velocities are measured with respect to the redshift of the BCGs as given in Table~\ref{tab:sample}. 

Extinction corrections were derived from the \ha/H$\beta$ flux ratio for the spectra pixels where \ha\, and H$\beta$ are detected above a surface brightness limit of $\scinot{2}{-17}$\surfb\, (equivalent to a signal-to-noise ratio of 2 to 3 in the data cubes).
We assume an intrinsic flux ratio H$\alpha$/H$\beta$ = 3.1. This value is appropriate for active galaxies, including LINERs \citep{Veilleux:1987p889}, in which there is a collisional-excitation component in the H$\alpha$ flux. It is slightly higher than the classical Case B recombination value of 2.86. 
Integrated emission-line fluxes are corrected for extinction using the reddening law of \citet{Osterbrock:1989p9491} and we adopt a value of $R = A_V / E(B-V) = 3.1$. 
Where the H$\beta$ emission is faint or undetected, we adopt a constant extinction to correct the line fluxes. This is estimated by measuring the integrated H$\alpha$ and H$\beta$ emission-line fluxes over a region of the inner, bright filaments in each galaxy (excluding the nuclear region) where the signal-to-noise ratio of both hydrogen lines is greater than five, and obtaining a mean extinction value from the integrated flux ratio.  The average $A_V$ values estimated and used in this way are listed in the third column of Table~\ref{tab:totalflux}. The estimated extinction contributions from the Galaxy, from the dust maps of \citet{Schlegel:1998p10426}, are also listed in Table~\ref{tab:totalflux} for each of the galaxies.

In \abell, although the emission-line surface brightnesses are strongest in the nuclear region, the line profile shapes complicate the extinction estimate from the H$\alpha$ and H$\beta$ fluxes. There is insufficient signal in the H$\beta$ emission to fit the multiple kinematic components that are seen in the strong H$\alpha$ and [\ion{N}{2}] emission lines. In particular, in the galaxy core, a low-amplitude but very broad emission-line component provides a good fit to the wings of the strong line profiles, and contains a significant amount of flux. We were unable to fit this component in the fainter H$\beta$ line profiles, and as a result would overestimate the values of the hydrogen flux ratio in the core. For this reason, we apply the mean extinction level from the filaments ($A_V=0.4$), as explained above, to the fluxes in pixels within a radius of 3\arcsec\, in the \abell\, data cubes.

Uncertainties in the relative line-flux measurements were estimated using the noise properties of the spectra after continuum subtraction. For a given emission line, the 1$\sigma$ error in the flux density per pixel was taken to be the standard deviation in a line-free region adjacent to the feature. These were combined in quadrature to form the uncertainties in the integrated line flux.

\section{Results} \label{sec:results}

\subsection{Integrated spectral properties}

Figure~\ref{fig:intspec} presents spectra integrated over a circular 5\arcsec-diameter aperture at the peak of the galaxy continuum emission for the three galaxies. 
In each case the galaxy continuum dominates the spectrum and is typical of evolved stellar populations in the giant elliptical galaxies. The three spectra show strong emission features that are characteristic of LINER emission. Similar features are detected in each spectrum.
After removing the stellar continuum component, the following emission features are detected in each galaxy:  forbidden emission lines [\ion{O}{2}]\,$\lambda\lambda$3726, 3729 (unresolved), [\ion{O}{3}]\,$\lambda\lambda$4959, 5007, [\ion{N}{1}]\,$\lambda\lambda$5198, 5200 (unresolved), [\ion{O}{1}]\,$\lambda\lambda$6300, 6363, [\ion{N}{2}]\,$\lambda\lambda$6548, 6583, and [\ion{S}{2}]\,$\lambda\lambda$6716, 6731, and Balmer hydrogen recombination lines H$\gamma$, H$\beta$, and H$\alpha$. There is also a marginal detection of [\ion{Ne}{3}]\,$\lambda$3967 in the nuclear spectrum of \abell, and of He I $\lambda$5876 in the outer parts of the nebulae of \abell\, and \zw.

In each galaxy, and as seen in other BCG filament systems \citep[e.g., NGC4696;][]{Farage:2010p11714}, the morphologies of all the optical emission lines are similar, though detected over different extents depending on the strength of the line. The kinematics are also consistent with being the same in all the detected lines \citep[e.g., as in Figure~4 of][]{Farage:2010p11714}. Everywhere that the density-sensitive [\ion{S}{2}]\,$\lambda\lambda$6716, 6731 doublet can be measured, the line flux ratios are consistent with densities of the [\ion{S}{2}]-emitting gas in the low-density limit: that is, $n_e \lesssim 200$\,cm$^{-3}$.

The total \ha\, and [\ion{N}{2}]\,$\lambda$6583 emission-line fluxes measured from the data cubes and the inferred luminosities are listed in Table~\ref{tab:totalflux}.
We can estimate an upper limit of the mass of emitting ionized hydrogen in the filaments from the H$\alpha$ emission-line luminosity, $L({\rm H}\alpha)$, following \citet{Osterbrock:1989p9491}:
\begin{equation}
M_{\rm H^{+}} = \frac{L({\rm H}\alpha)\,m_{p}}{n_{e}\,\alpha_{\rm{H}\alpha}^{\rm eff}\,h\,\nu_{{\rm H}\alpha}}. \label{eq:massH}
\end{equation}
The derived masses are listed in Table~\ref{tab:totalflux}, assuming an electron density of $n_e<200$\,cm$^{-3}$.
We adopt an effective recombination coefficient of $\alpha_{\rm{H}\alpha}^{\rm eff} = \scinot{1.17}{-13}$\,cm$^{3}$\,s$^{-1}$ \citep{Osterbrock:1989p9491}. This is appropriate for Case B emission at a temperature of  $T=10^{4}$\,K. The ionized hydrogen masses are low compared to the mass of molecular material inferred in the filaments. For example, \citet{Edge:2001p4022} estimate a molecular gas mass of $\sim\scinot{4}{9}$\msol\, in \zw, and find similarly large masses in a number of other cooling-flow BCGs. This is consistent with the line emission arising in thin layers of ionized gas surrounding dense, cool molecular cores that contain most of the mass in the filaments \citep[e.g.,][]{Salome:2011p13108,Johnstone:2007p2172}. 


The integrated [\ion{N}{2}]/H$\alpha$ flux ratios are shown in Table~\ref{tab:totalflux} with the total luminosities. We note that our data for \abell, \zw, \sersic, and NGC~4696 show good agreement with the  correlation observed by \citet{Crawford:1999p1097} of decreasing \NIIHa\, flux ratio with total \ha\, luminosity in BCG nebulae. 
Table~\ref{tab:linefluxes} presents line flux measurements for the emission features detected in the spectra of the three galaxies, from a 5\arcsec-diameter aperture centered on the galaxy nucleus and also in an off-nuclear position in the filaments. For the latter, a bright region of the filaments was selected that is sufficiently far from the nucleus to avoid any significant contribution to the emission from the AGN. 
The high [\ion{O}{3}]/H$\beta$ ratio in the nuclear spectrum of \abell\, is likely to be the result of radiation from the AGN, but, as expected, all of the line ratios are consistent with a LINER-like emission spectrum. The optical line ratios in the nebulae and the integrated emission properties of these objects are typical of BCG systems.



The morphology and kinematics of the optical line emission detected in the observations of each of the three galaxies are discussed individually in the following sections.

\subsection{Abell 3581} \label{sec:res:abell}

\begin{figure}
\epsscale{1.}\begin{center}
\plotone{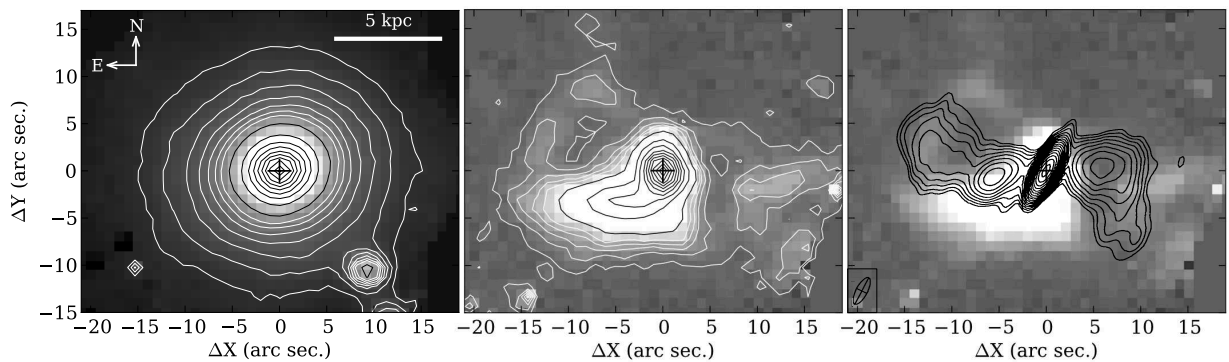}\end{center}
\caption{\textit{Left:} \abell\, continuum image, centered at 8000\,\AA\, in the galaxy rest frame. A foreground star is seen in the southwest corner of the image.  \textit{Center:} Map of the [\ion{N}{2}]\,$\lambda$6583 line emission surrounding the BCG. The cross marker indicates the location of the continuum peak. \textit{Right:} Contours of 1.4\,GHz radio emission from \citet{Johnstone:1998p7726}, overlaid on the greyscale emission-line image from our data. The radio beam size is shown in the bottom-left corner of the panel.}\label{fig:a3581contmap}
\end{figure}

Figure~\ref{fig:a3581contmap} shows integrated continuum and [\ion{N}{2}]\,$\lambda$6583 emission-line images of the BCG in \abell\, from the WiFeS data. The line emission extends over $15-20$\arcsec\, ($7-9$\,kpc) towards the east and west of the galaxy nucleus. Within the angular resolution of the observations, the peak emission-line surface brightness is coincident with the galaxy nucleus. There is unresolved line emission in the galaxy core with a peak surface brightness $\sim2-3$ times higher than in the extended emission regions.

The right panel of Figure~\ref{fig:a3581contmap} compares the 1.4\,GHz radio image of this source with the optical line emission map.  The line emission has a similar orientation and extent on the sky as the extended radio emission that is associated with the galaxy. The radio lobes are surrounded by filaments of line emission and the brightest regions of radio emission largely coincide with regions where the optical line emission is faint or undetected. Based on the morphology of the line emission, there appears to be a close association between the radio lobes and line-emitting gas in this galaxy.

The kinematics of the nebula are complex. There are steep gradients in the line-of-sight velocity distribution. There are also several regions where the \ha\, and [\ion{N}{2}] emission-line profiles are clearly composed of multiple kinematic components. In particular, within a radius of approximately $2-3$\arcsec\, of the galaxy core, the line profiles show  symmetric blue and red wings. These are fit well by a second broad, low-amplitude Gaussian component, with a velocity dispersion of  $750-800$\kms\, and centered at the same velocity as the strong, narrow line component. This component of the line profiles may represent emission from the broad-line region (BLR) of the active nucleus.
Within a radius of 3\arcsec\, of the nucleus, the profile fits to the narrow and broad velocity components of the lines show a uniform gradient in the line-of-sight velocity across the nucleus. The velocity amplitude is approximately $\pm80$\kms\, in this region.

\begin{figure}
\epsscale{0.9}
\plotone{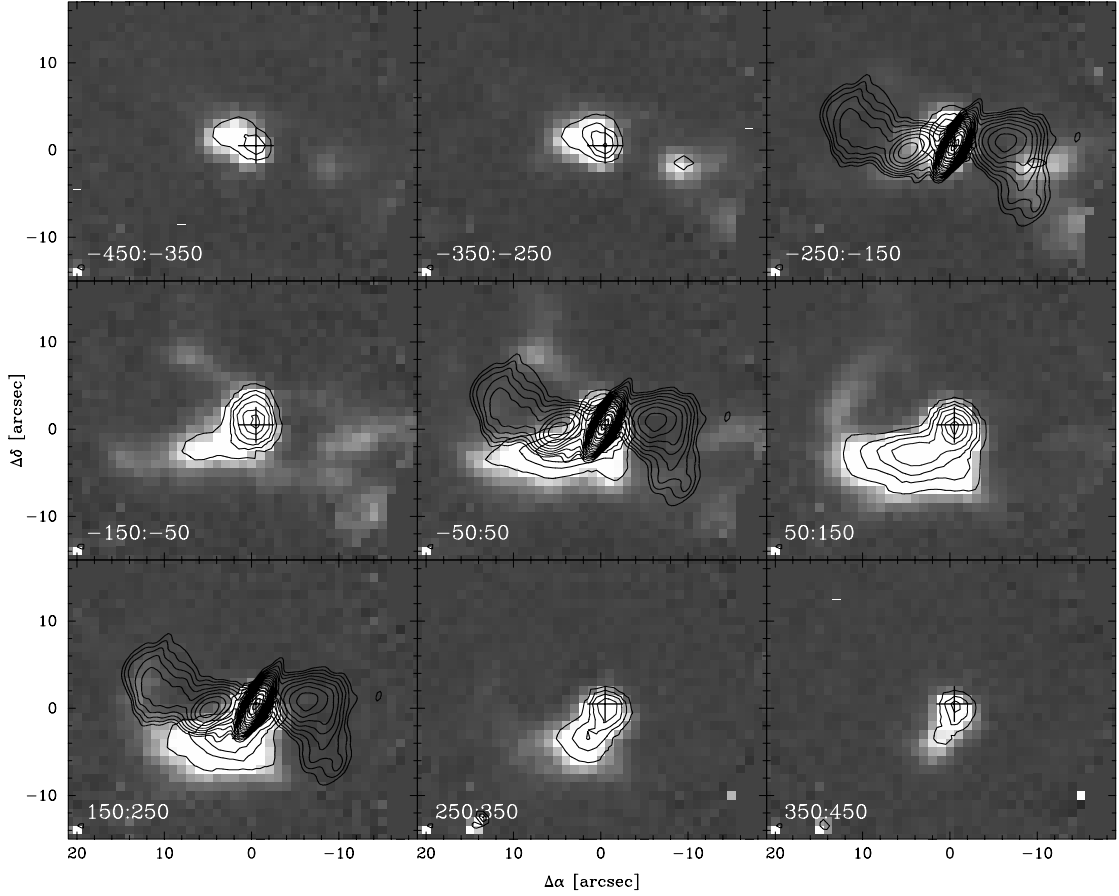}
\caption{Velocity channel maps across the [\ion{N}{2}]$\lambda$6583 emission line. The velocity range in \kms\, is shown at the lower left of each panel. The cross marker indicates the position of the galaxy nucleus. Contours of the 1.4\,GHz emission radio map from \citet{Johnstone:1998p7726} are overlaid on several panels (black contours).}\label{fig:a3581channels}
\end{figure}

The kinematic features and the overall distribution of velocity in the emission-line regions are most clearly illustrated using velocity channel maps. Figure~\ref{fig:a3581channels} presents velocity slices across the brightest emission line in the spectrum, [\ion{N}{2}]\,$\lambda$6583. The panels in this figure are formed by selecting bins of constant velocity width and summing the wavelength slices in each bin to form an image. Each image shows the spatial distribution of gas in the nebula with line-of-sight velocities in a given 100\kms\, range, with respect to the redshifted wavelength of the line. The 1.4\,GHz radio map is also overlaid on several of the panels for comparison.

We measure a large range of line-of-sight velocities in the nebula: $\Delta v\sim900$\kms. The highest velocities, above $\pm300$\kms, occur close to the center of the nebula, within $\sim5$\arcsec\, ($\sim2.3$\,kpc) of the nucleus. There is a cloud of high-velocity blueshifted emission northeast of the galaxy core where the line-of-sight velocities peak at $>500$\kms. This appears as an additional blue-shifted peak or wing in the spectral line profiles from the region. 
At low velocities, between approximately $\pm100$\kms, the emission nebula consists of roughly V-shaped arms oriented close to the radio axis. This structure is reminiscent of shells or filaments of gas surrounding bi-polar cavities. 

The bright spiral filament to the south of the eastern radio lobe produces redshifted emission with velocities increasing toward the nucleus and reaching a peak of $\sim450$\kms\, within several arcseconds of the core.
This filament encircles the southern edge of the radio lobe, and the knot at the peak of the radio emission in this lobe lies just interior to the inner edge of the filament. 
Again, the alignment suggests that the shape of the filament is related to that of the radio lobe. One possibility is that the emission-line gas may be uplifted by the expansion of the radio bubbles. Since we do not have information about the orientation of the radio lobes or emission nebula, we can not distinguish inflow from outflow on the basis of the line-of-sight velocities. However, if the motion of the emission-line gas is largely driven by the propagation of the radio outflows, the velocities are directed outward from the galaxy center. This is consistent with  the mostly redshifted emission seen in the eastern region of the nebula and blueshifted velocities of the filaments to the west. 
The highest velocity gas in the vicinity of the western radio lobe occurs in a bright clump of emission located where the radio lobe changes direction and extends to the south. This may be evidence of an interaction of this type occurring there. 

However, in this scenario it is difficult to explain the cloud of high-velocity blueshifted emission to the northeast of the galaxy nucleus. The emission lies along the northern edge of the radio source.
There is emission-line gas with high, but oppositely-directed, velocities located roughly symmetrically to the northeast (blueshifted) and southeast (redshifted) of the nucleus; seen by comparing the $|v|=250-350$\kms\, velocity slices in Figure~\ref{fig:a3581channels}. This is not consistent with an outflow driving the kinematics in (both) these structures.

It is possible that the material in the spiral-shaped filament to the southeast of the galaxy is flowing inward on an orbit-like trajectory. This is reminiscent of the motion of gas observed in the main filament of NGC~4696 \citep{Farage:2010p11714}.
In this case the high-velocity blueshifted emission to the northeast of the nucleus may be gas on the approaching side of this spiral orbit, after it passes behind the nucleus.
The maximum velocity of $\sim450$\kms\, occurs in the blueshifted clump approximately 3.5\arcsec\, (1.6\,kpc) from the position of the nucleus. 
Assuming circular orbital motions implies an enclosed mass within this radius of $\sim\scinot{5.9}{10}$\msol. \citet{Rafferty:2006p2278} estimate that the bulge mass of this galaxy is $\scinot{(5.7\pm0.5)}{11}$\msol\, so this interpretation implies that approximately 10\% of the bulge mass is located within a radius of 1.6\,kpc. 
\citet{Govoni:2000p1517} fit a de Vaucouleur's profile to the $R$-band surface brightness of the galaxy and obtained an effective radius of $R_e=29$\arcsec\, (13.2\,kpc).
A de-projection of this profile assuming spherical symmetry, as performed by \citet{Young:1976p12551}, indicates that the fraction of the total luminosity found within a radius of 3.5\arcsec\, is $\sim5$\%.
This implies that a similar mass fraction lies within this radius, assuming a constant mass-to-light ratio throughout the galaxy. This is similar to the result derived from the kinematics and so the measured velocities plausibly arise from material on an orbital trajectory and could represent infall. 

In the outer regions of the filaments where the single Gaussian line-profile fits are unambiguous, the gas velocity dispersion is relatively uniform in the nebula. Several ridges of broader lines are consistent with being the result of beam-smearing, where the line-of-sight velocity gradient is high and pixels sample a range of velocities, causing broadening of the line profiles. The average velocity dispersion in the outer nebula is $\sim90\pm15$\kms.

The ratio of [\ion{N}{2}] to \ha\, flux is also quite uniform in the outer filaments, with values of $\sim1.0-1.4$. 
From the dual-component Gaussian fits, we find a slight increase in the \NIIHa\, ratio in the core: the narrow line components indicate a flux ratio of approximately 2.0 in the central pixels. 
The [\ion{O}{3}]/H$\beta$ line flux ratio shows a distinct increase at the nucleus, from values of $\sim0.5-1.0$ in the spiral filament to $\sim2.5$ in the nucleus. Both line flux ratio increases are attributable to a contribution to the ionizing spectrum from the radiation field of the AGN.

\subsection{2A 0335+096} \label{sec:res:zw}

\begin{figure}
\epsscale{1}
\plotone{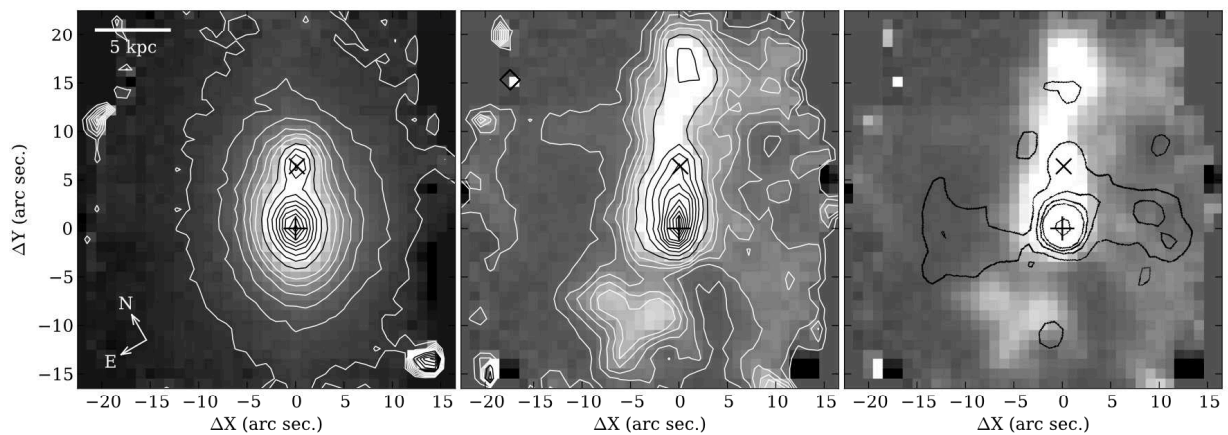}
\caption{\textit{Left:} \zw\, continuum image, centered at 8000\,\AA\, in the galaxy rest frame. \textit{Center:} Map of [\ion{N}{2}] line emission surrounding the BCG. \textit{Right:} Contours of 1.4\,GHz radio emission from \citet{Donahue:2007p11428} are overlaid on the greyscale emission-line image from our data. In each panel, the BCG continuum peak is indicated by the black `+' marker, and the position of the companion galaxy by the `$\times$' marker.}\label{fig:zw0335contmap}
\end{figure}

Figure~\ref{fig:zw0335contmap} shows a map of the [\ion{N}{2}]\,$\lambda$6583 emission detected in our observations of \zw\, and an image of the continuum emission from the red data cube. The neighboring cluster galaxy to the northwest of the BCG is indicated in the figure. 
Our integral-field data map the kinematic properties across nearly the full extent of the nebula that has been detected in narrowband imaging \citep{Romanishin:1988p11445}. The detected line emission extends over almost the entire 26\,kpc\,$\times\,$26\,kpc field of view.  

\begin{figure}
\epsscale{0.44}
\plotone{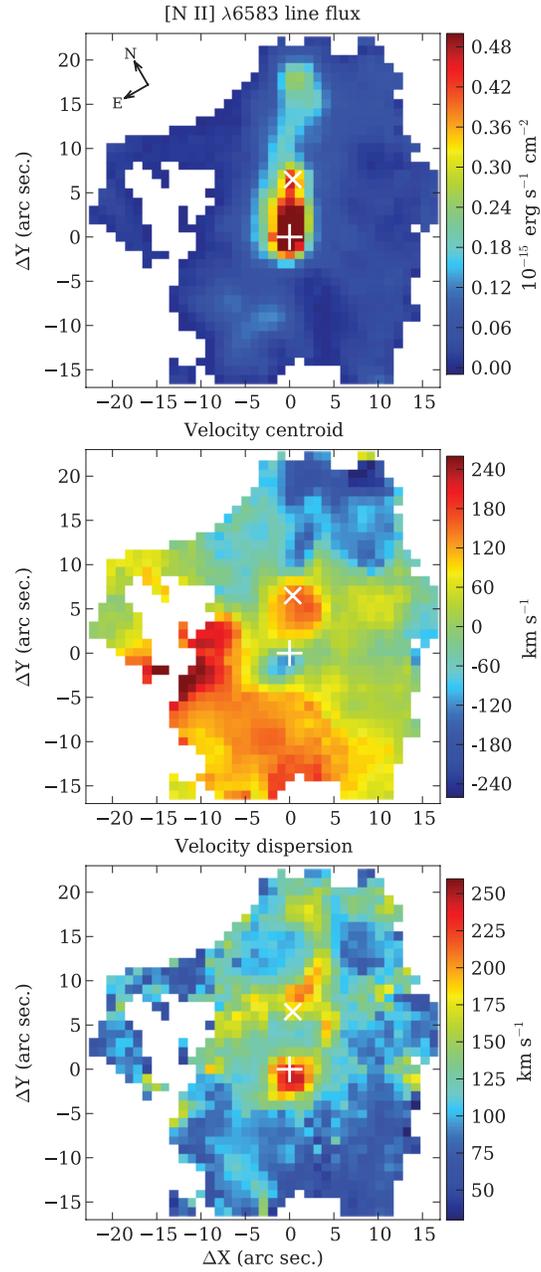}
\caption{Properties of single Gaussian profile fits to the [\ion{N}{2}] $\lambda$6583 emission line for \zw.}\label{fig:zw0335linemaps}
\end{figure}

\begin{figure}
\epsscale{0.8}
\plotone{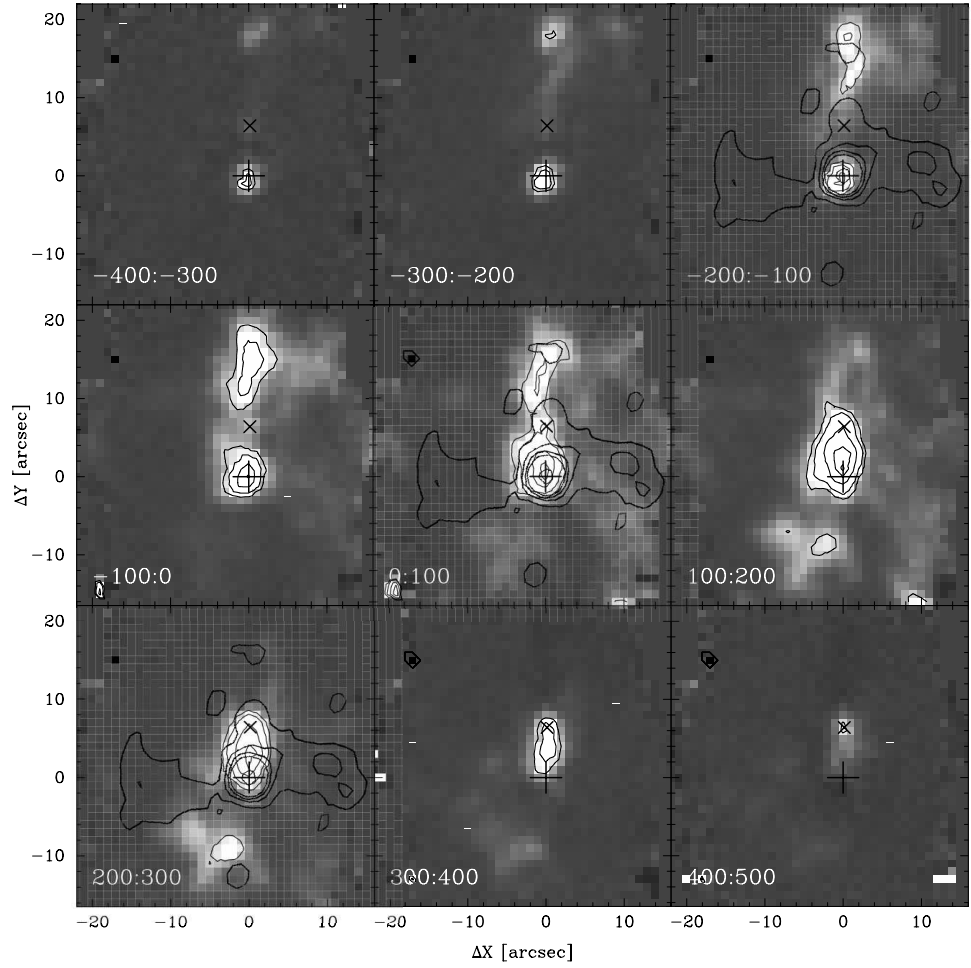}
\caption{Velocity channel maps across the [\ion{N}{2}] $\lambda$6583 emission line detected from \zw. The velocity range in \kms\, is indicated at the lower left of each panel. Contours of the 1.4\,GHz emission radio map from \citet{Donahue:2007p11428} are overlaid on several panels (black contours).}\label{fig:zw0335channels}
\end{figure}

Figure~\ref{fig:zw0335linemaps} shows the properties of single Gaussian line fits to the [\ion{N}{2}]\,$\lambda$6583 line. However, there are several regions in the nebula where the lines have more complex profiles than a single Gaussian kinematic component. This increases the uncertainties in the line flux, dispersions, and velocity centroids in those regions, and the increased velocity dispersions surrounding the BCG nucleus and to the north of the companion galaxy are the result of this kinematic structure.
The velocity channel maps in Figure~\ref{fig:zw0335channels} also illustrate the line-of-sight kinematic structure in the emission-line regions.


 
A complex system of filaments and clouds extends beyond the central BCG and companion galaxy. The brightest emission is confined to the bar structure described by \citet{Romanishin:1988p11445}, which extends across the field of view is oriented at a position angle of approximately 140\degree. 
The velocity map in Figure~\ref{fig:zw0335linemaps} clearly demonstrates the kinematic structure that was seen in this bar in early spectroscopic studies \citep{Gelderman:1996p11337,Donahue:2007p11428}, and reveals the distribution over the entire emission region. 
We confirm that the outer parts of the nebula appear to be rotating about the center of the BCG, as proposed by \citet{Donahue:2007p11428}. The rotation has a peak circular velocity of $\sim200$\kms\, at a radius of $r\sim15$\arcsec\, (10\,kpc). The velocity gradient across the inner region of the nebula between the two galaxies is superimposed on this large-scale velocity field and counter-rotates with respect to the outer bar with an amplitude of $\Delta v\sim130$\kms. The line-of-sight velocities associated with the relative motion of the BCG and companion galaxy share this counter-rotational sense of motion with the inner disk-like structure around the nucleus.

The right panel of Figure~\ref{fig:zw0335contmap} compares the morphology of the 1.4\,GHz radio source associated with the galaxy to the optical line map from our data. 
There are faint filaments of line emission that extend to the northeast and southwest of the galaxy nucleus and encircle the radio lobes. \citet{Donahue:2007p11428} highlight two circular regions as being delineated by line emission and coinciding approximately with the radio lobes. These can be seen in the central panel of Figure~\ref{fig:zw0335contmap} where the contours outline roughly circular depressions in the emission-line surface brightness to the northeast and west of the galaxy. 
The line-of-sight velocities in these structures are low, which could indicate that they have only low inclination from the plane of the sky and are oriented close to the rotation axis of the emission-line gas.

The ratio of \NIIHa\, is approximately uniform along the extent of the bar of bright line emission, including between the BCG and companion galaxy, with a value of  $1.0\pm0.1$. The measured ratios of flux in the components of the [\ion{S}{2}]\,$\lambda\lambda$6716, 6731 doublet are also consistent with being uniform over the brightest emission regions where the signal-to-noise ratio is high, and indicate electron densities of $\lesssim200$\,cm$^{-1}$.

The [\ion{O}{3}]\,$\lambda$5007 line is detected only from the inner region of the nebula between the nuclei of the two galaxies, but is extended along the axis of the bar and shows a similar velocity structure as the brighter H$\alpha$ and [\ion{N}{2}] lines, consistent with arising in the same extended filamentary gas.
The [\ion{O}{3}]/H$\beta$ ratio increases at the nuclei of both the BCG and companion galaxy, with values of $\sim0.7\pm0.1$, but is low ($\lesssim0.2$) in the outer regions of the bar, where H$\beta$ is detected and [\ion{O}{3}] is not.

\subsection{Sersic 159-03}  \label{sec:res:sersic}

\begin{figure}
\epsscale{0.85}
\plotone{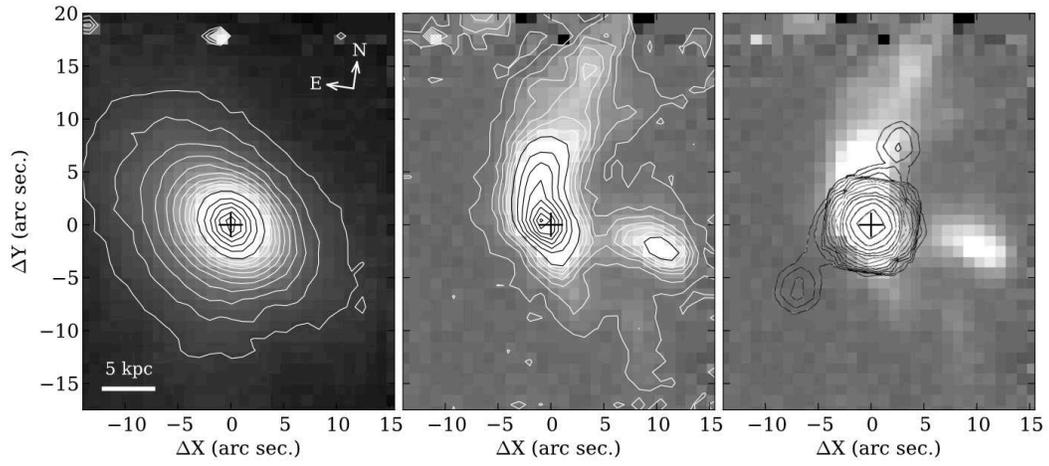}
\caption{\textit{Left:} Continuum image of \sersic\, centered at 8000\,\AA\, in the rest frame of the galaxy. \textit{Center:} Map of the [\ion{N}{2}]\, emission surrounding the BCG. \textit{Right:} Contours of 5\,GHz radio emission from \citet{Birzan:2008p6520} overlaid on the emission-line image from our data. The continuum peak is indicated by the black cross marker. }\label{fig:s159contmap}
\end{figure}

The extent of the optical line emission detected surrounding \sersic\, is illustrated in Figure~\ref{fig:s159contmap}, together with an image of the galaxy in continuum emission. 
We observe the inner 20\arcsec\, (21\,kpc) of the bright, curved northern filament, which extends beyond the field of view of our data cube. The surface brightness in the filament increases toward the  galaxy nucleus and the peak line surface brightness is offset by $\sim1$\arcsec\, to the east of the galaxy continuum peak. 

\begin{figure}
\epsscale{0.4}\centering
\plotone{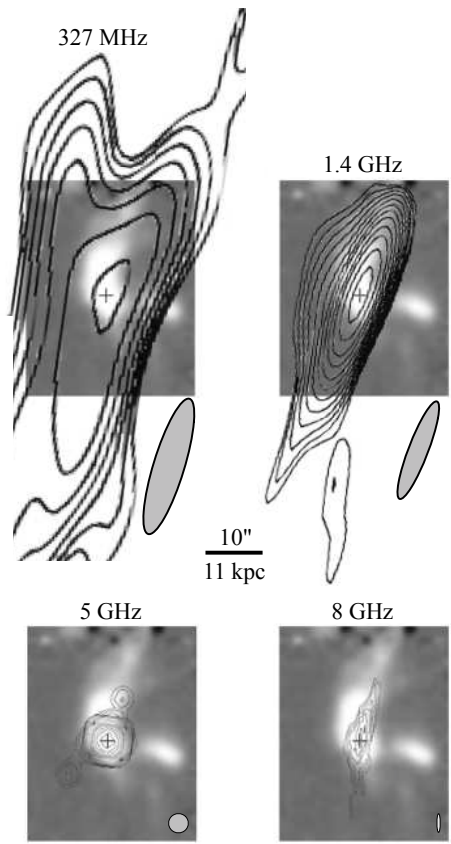} 
\caption{Contours of the radio maps measured by \citet{Birzan:2004p8008} from observations of \sersic\, at 327\,MHz (top left), 1.4\,GHz (top right), 5\,GHz (lower left), and 8\,GHz (lower right), overlaid on a smoothed image of the [\ion{N}{2}] emission from our data. The relevant radio beam sizes are indicated by the grey ellipses to the lower right of each map.}\label{fig:s159radio}
\end{figure}

 The right panel of Figure~\ref{fig:s159contmap} shows the 5\,GHz radio map of \sersic\, overlaid on the [\ion{N}{2}] image from our data. There is no clear correspondence between the morphology of the optical line-emitting filaments and the radio emission, to the degree that the angular resolution of both observations reveals. However, as illustrated in Figure~\ref{fig:s159radio}, the northern filament is extended along approximately the same position angle ($160-170$\degr) as the emission detected at radio frequencies from $0.3$ to $8.5$\,GHz. The southern radio lobe is not associated with any detected optical emission and there are several filaments that are not located close to any radio structures. 

\begin{figure}
\epsscale{0.8}
\plotone{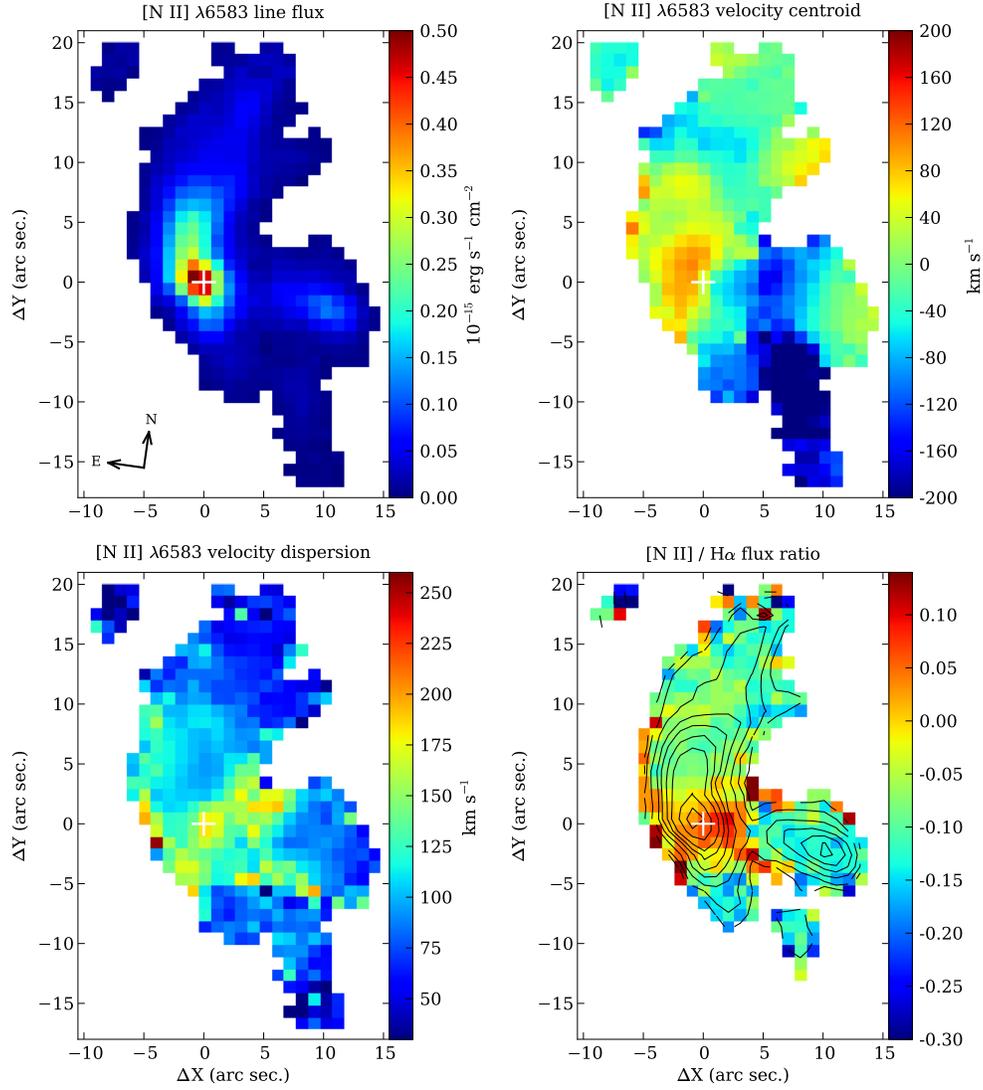}
\caption{Maps of the line properties derived from single Gaussian profile fits to the [\ion{N}{2}] $\lambda$6583 emission line, and, at lower-left, the [\ion{N}{2}]/H$\alpha$ line flux ratio for \sersic.}\label{fig:s159linemaps}
\end{figure}
\begin{figure}
\epsscale{0.9}
\plotone{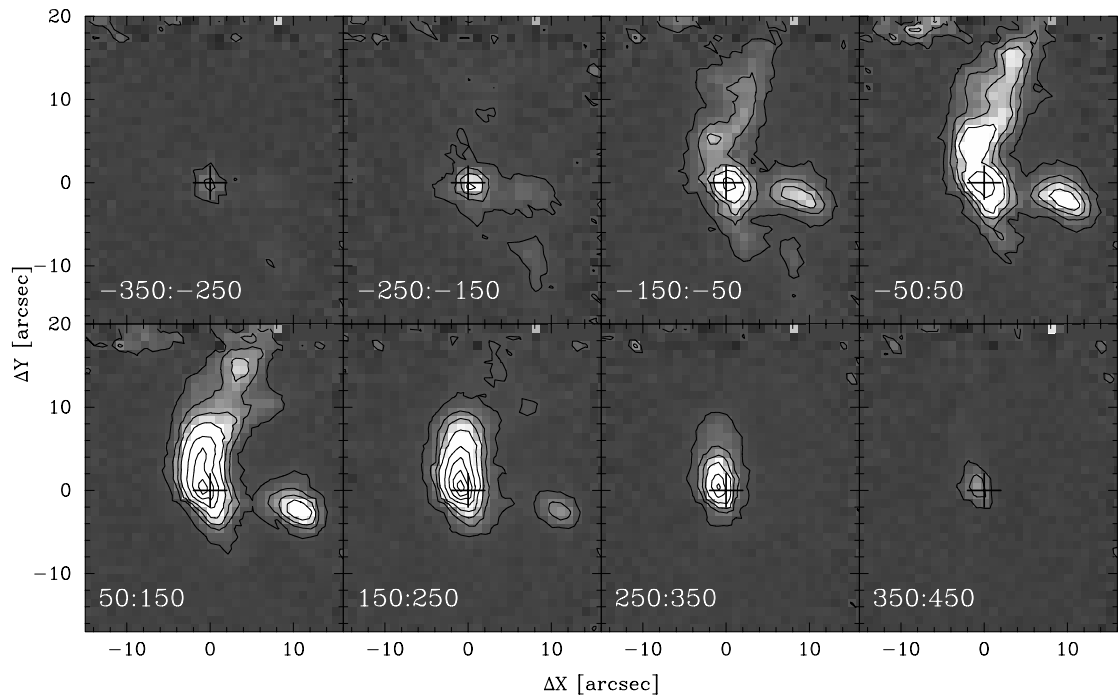}
\caption{Velocity channel maps across the [\ion{N}{2}] $\lambda$6583 emission line in \sersic. The velocity range in \kms\, is indicated at the lower left of each panel.}\label{fig:s159channels}
\end{figure}

A single Gaussian function provides a good fit to the emission-line profiles everywhere in the data cubes; we do not resolve more complex kinematic components. Figure~\ref{fig:s159linemaps} shows maps of the fluxes, velocity centroids, and velocity dispersions derived from fits to the [\ion{N}{2}]\,$\lambda$6583 line. Velocity channel maps across this emission line are shown in Figure~\ref{fig:s159channels}.



The line-of-sight centroid velocities of the detected emission vary over $\sim330$\kms.  The velocity centroid map and channel maps show a velocity shear across the nucleus of the galaxy that resembles rotation of the gas in the nucleus, with an amplitude of approximately $\pm60$\kms, over a radius of $\sim2$\arcsec\, ($\sim2$\,kpc). This provides an estimate of $\sim10^{9}$\msol\, for the enclosed mass, assuming circular rotation. This is consistent with the black-hole mass estimate of \scinot{(4\pm1)}{8}\msol\, by \citet{Rafferty:2006p2278}, based on the central velocity dispersion. The axis about which the implied gas rotation occurs has a position angle on the sky of $\sim160\degr$.

The nebula shows high central line-of-sight velocities and velocity dispersions, and decreasing velocities outward along the filaments. The velocity from the emission in the northern filament decreases from $\sim90$\kms\, within $1-2$\arcsec\, of the nucleus, to approximately zero at the outer end of the filament. 	
The northern filament diverges into two ridges beyond $5-10$\arcsec\, from the galaxy, and the line-of-sight velocities in the southern component are redshifted by $\sim50$\kms\, relative to the north region. This is evidence of more complex structures that are not well resolved in our data.  The kinematics do not allow us to distinguish between infall and outflow in these filaments, since the inclinations are not known. However, the line-of-sight velocities of the inner part of the northern filament are consistent with a smooth transition between the velocities in the filament and the velocity shear across the nucleus.

The line-of-sight velocity dispersion in the emission-line gas decreases towards the outer regions of the filaments, from $\sim100$\kms\, in the bright inner region of the northern filament, to $\sim60-80$\kms\, at the outermost edges of the nebula. This could be the result of increased turbulence in the inner nebula, or the superposition of filaments with more complex velocity structure that broadens the measured line widths. 

The line flux ratio of \NIIHa\, peaks in the galaxy core and is also generally lower in the outer regions of the nebula, as was observed by \citet{Jaffe:2005p2500} and \citet{Oonk:2010p11713}. The flux-ratio values are also lower in the western filament (\NIIHa$\,\sim0.7$) than in the filament to the north (\NIIHa$\,\sim0.8$). Figure~\ref{fig:s159linemaps} shows the [\ion{N}{2}]/\ha\, flux-ratio map of \sersic\, from our data.

\section{Discussion}\label{sec:disc}

\subsection{Interactions between the radio lobes, ICM, and optical nebulae}\label{sec:disc:inter}

Interactions between the line-emitting nebulae,  the cooling ICM, and radio outflows that heat the environment provide clues to how the filaments are linked to feedback processes in cluster cores.
For example, several BCGs have been observed where the filamentary structures largely avoid the radio lobes, or the radio emission coincides with regions where there is no, or lower, surface-brightness line emission \citep[e.g.,][]{Sabra:2000p2757,ODea:2004p6715,Oonk:2011p12544}. 
\citet{Sabra:2000p2757} suggest that this is evidence either that the radio lobes displace and entrain the nebular gas, or that the relatively dense, line-emitting gas confines the radio lobes and channels the outflow through the regions of lowest density in the inter-galactic medium.
This raises the question of whether the principal effect is of the radio jet and lobes influencing the morphology and kinematics of the emission-line gas or vice versa. These issues are linked to whether the emission-line gas is infalling and primarily a symptom of gas cooling and accretion, or outflowing and related to reheating facilitated by the AGN and radio jet. 
Of our sample, this type of interaction is particularly notable in \abell, though it may have occurred to some degree in the other two systems.

In \abell,  the kinematics and the morphologies of the ionized gas and the radio emission suggest that the two components are interacting. The shape of the optical emission regions, forming bipolar structures that surround the radio lobes, suggests a component of outflow in the nebula that is generated by the expanding lobes. Given the coincidence between several clumps of line-emitting gas and bends in the orientation of the radio lobes, we conclude that there is significant interaction between the structures. The gas with low line-of-sight velocities that surrounds the outer sides and edges of the radio bubbles has oppositely-directed velocities at the east and west ends of the nebula, and so is probably gas that is being displaced and pushed outwards from the galaxy by the expansion of the lobes. 
The distortion of the radio lobes implies that the radio structures are also perturbed by the presence of cold, dense gas in their path. 
We suggest it is most likely that the radio lobes and cooling optical nebula mutually affect one another as the lobes expand, since there is evidence for both possible inflow and outflow motions in the nebula; the radio lobes are driving out gas in the outer filaments, while there may be gas infalling along the inner, bright spiral filament.

The structure of the radio source, the X-ray cavities, and the morphology and kinematics that we have observed in the optical nebula reveal the system to be in a well-developed stage of AGN outflow and heating. The radio source is not overly luminous, but there is significant interaction between the radio outflows and the surrounding environment. There is evidence for infall and accretion occurring from the line-emitting nebulae, which may be remnant flows that are feeding the current nuclear activity. It appears that we are observing a phase of feedback activity in which AGN heating balances and possibly dominates the cooling of gas from the ICM.

In \zw, there is no indication of a strong interaction between the optical and radio-emitting structures. The filaments to the east and west have some spatial association with the radio bubbles: the lobes are coincident with low surface-brightness regions of the nebula, and there are faint filaments that outline the edges of these cavities. It is possible that some of this emitting gas has been entrained by the expansion of the lobes to their current positions. However, most of the emission-line gas in the system appears to be unrelated to the weak radio lobes in this galaxy. In particular, the large-scale bar of ionized gas that dominates the structure is not associated with any detected radio emission.  

\citet{Sanders:2009p11794} estimate that the total enthalpy associated with the five X-ray cavities that they detect in \zw\, is approximately \scinot{5}{59}\,erg. The cooling luminosity from within the 120\,kpc cooling radius of the cluster is approximately \scinot{3}{44}\lum\, \citep{Birzan:2004p8008,Rafferty:2006p2278}.
Therefore, they note that if the average cooling rate is uniform over time, the estimated enthalpy of the cavities would be sufficient to balance the ICM cooling for only $\sim\scinot{5}{7}$\,yr.  The current radio source is estimated to have an age similar to this timescale \citep{Donahue:2007p11428,Sarazin:1995p6240}, and estimates of the average time between outbursts in cooling-flow clusters are closer to $\sim10^{8}$\,yr \citep{Birzan:2004p8008}. 
Given the evidence for a number of episodes of AGN activity that have produced the observed cavities, the heating has clearly occurred intermittently over a significantly longer timescale than \scinot{5}{7}\,yr: at least the length of several AGN outburst cycles. Thus, for a significant fraction of the time the core must have been dominated by cooling, and it currently appears to be in such a phase of strong cooling and accretion activity. 

 The structure and kinematics of the optical emission also provide evidence for material forming a rotating disk-like configuration in the outer nebula, possibly being accreted from the cooling gas of the ICM.
 There is a large mass of cool material associated with the extended filaments.  CO emission-line measurements imply that there is $\sim\scinot{4}{9}$\,\msol\, of molecular material within 30\,kpc of the BCG \citep{Edge:2001p4022,Edge:2003p6404}. 
 
It appears that cooling activity dominates in the core of \zw\, and the energy output from the radio source is insufficient to balance the ICM cooling at the current time. The radio outflows appear to have relatively little influence on the emission-line nebula, and we suggest that we are observing a system in a strongly cooling phase of feedback. 

The data for \sersic\, are more difficult to interpret.  We can not discount an interaction between the optical filaments and radio outflows, but do not see clear evidence for it in our data. 
\citet{Werner:2011p12464} argue that there has been a strong interaction between the AGN outflows and optical-line-emitting filaments. They propose that the material in the north and west filaments has been uplifted from the central galaxy by an expanding radio outflow, and assisted in reaching the current position relative to the galaxy by the southward motion of the BCG through the cluster atmosphere. The BCG is located approximately 8\arcsec\,(8.6\,kpc) south of the cluster X-ray emission peak \citep{Jaffe:2005p2500} and has a redshift of $\sim-500$\kms\, relative to the cluster center. This interpretation is supported by metallicities and temperatures measured in the clumpy X-ray emission aligned with the optical filament that are lower than in the surrounding ICM. In addition, they find that the core of the cluster is largely devoid of the densest X-ray emitting ICM within $\sim8$\,kpc, which has presumably been evacuated by the AGN outflows.

However, there is not a strong relationship between the distribution of the optical line emission and the radio lobe structures that are observed. The western filament, in particular, is quite separate from the radio emission and seems unlikely to have been affected by the radio outflow activity. 
Our data indicate that the kinematics of the northern filament are equally consistent with inflowing velocities along the structure as with outflow, and we suggest that the material in the filament could be part of an accretion flow into the central galaxy. The kinematics of the material in the filament are consistent with a flow that smoothly joins the rotation seen in gas close to the core. The filament material may have been uplifted by outflows associated with an earlier episode of AGN activity, have since had time to cool, and is now falling into the central galaxy, potentially fueling renewed AGN activity represented by the high-frequency inner radio lobes. 

It is clear that the central AGN has had a significant effect on the surrounding ICM. So although the kinematics of the optical emission-line regions do not assist in a definitive interpretation of the behavior of the material in the nebula,  it is likely that we are observing this system in an AGN-dominated phase of feedback. 

To summarise, in \abell\, the structure of the nebula is closely related to the radio outflows and it seems clear that AGN feedback at least offsets most of the core cluster cooling in the observed phase of activity. In \zw\, there is evidence for minor interactions between the emission-line gas and radio lobes, but predominantly the nebula is not affected by the AGN output, and accretion and cooling activity dominates. \sersic\, shows evidence for AGN activity that has had a strong influence on the cluster core, and may have affected the cooling material in the nebula, particularly the northern filaments, in the past. 
The optical spectroscopic data provide information about the state of the warm gas in the cluster core that can reinforce inferences drawn from the X-ray and radio data, but do not necessarily provide an indication of the state of feedback in the absence of information about the hot cluster gas and/or radio activity. 

\subsection{Core gas rotation and radio-axis alignment}\label{sec:disc:rotation}
In all three BCGs there is evidence for rotation in the emission-line gas close to the core of the galaxies. Such motion could represent accretion flows in the nuclear regions that provide a channel for feeding material to the central black hole. Furthermore, the axis of rotation within a few kiloparsecs of the nucleus is, in all three cases, approximately aligned with the outflow axes of the radio lobes associated with the galaxies. 

In \abell\, there is a gradient in the line centroid velocities derived from fits to the narrow emission lines in the galaxy core. This resembles a velocity field of rotation about an axis oriented at ${\rm PA}\sim115$\degr\, and may be evidence for an accretion disk in the core. From the 1.4\,GHz radio map, the orientation of the radio lobes in the inner part of the jets is approximately $95$\degr\, on the sky.
If the bright spiral filament in the nebula of \abell\, does consist of material that is currently falling into the central galaxy, as suggested in Section~\ref{sec:res:abell}, it may represent a remaining component of the accretion flows in the galaxy core that contribute to the current nuclear activity. The orientation of the filament is not well resolved close to the center of the galaxy, but the structure bends towards the north in the inner regions and the data are consistent with a trajectory that is close to perpendicular to the radio axis as it approaches the nucleus. 

In \zw, \citet{Wilman:2011p13215} have mapped the velocities of the H$_{2}$ emission that forms a dumbbell-shaped structure within the central 2\,kpc. There is a velocity shear across this emission region that could represent rotation of gas in an edge-on torus or disk. The axis of rotation has a position angle on the sky of $\sim55$\degr, which again is close to the orientation of the radio lobes (${\rm PA}\sim60$\degr). 
Material from this structure could feed the galaxy nucleus and the orientation of the radio outflow axis may have been established by the equatorial accretion flows from this disk of gas in the galaxy core. 

If the bar-shaped structure and rotational motion of the outer nebula represent gas settling into a large-scale disk, it is likely to be a manifestation of cooling and accretion activity on large-scales over the cooling region of the cluster core,  occurring during a cooling-dominated phase of the feedback cycle. The opposite sense of the rotation in the outer part of the nebula compared to the motion of gas in the nucleus may indicate that this activity is decoupled from the phase of accretion occurring in the core. This decoupling represents a break in the accretion of material into the inner part of the nebula that likely results in an interruption of AGN fueling and a therefore a lapse in the feedback process. This may be what allows the cluster core to be largely cooling-dominated at the observed period. 

In \sersic\, there is a continuous gradient in line-of-sight velocity across the nucleus that resembles rotation about an axis with ${\rm PA}\sim160$\degr. This axis is close to the axis of the extended radio lobes at ${\rm PA}\sim170$\degr. \citet{Oonk:2010p11713} also note that there is evidence from their near-infrared, integral-field spectroscopy that the \ion{H}{2} and H$_2$-emitting gas within several kiloparsecs of the nucleus rotates about an axis parallel to the radio jet.

These relationships between the radio axes and the accretion flows in the cores suggest that the inflows and outflows are closely connected, as expected in the AGN feedback model. The timescales of the implied accretion processes provide a measure of the feasibility of these links.  
We can make estimates about the rate of gas accretion into the galaxy cores if we assume that the observed motions represent material in accretion flows that feed cooling material from the nebula into the nucleus, and that the bright filaments in \sersic\, and \abell\, trace infalling gas feeding these flows. 
The approximate accretion time through the filaments may be estimated from the average velocity $v$ of material along the filament and the distance $d_{\rm fil}$ along the structure into the core: $t_{\rm acc}\approx d_{\rm fil}/v$. The mass inflow rate in the filament is then  
\begin{equation} 
\dot{M}_{\rm in} = \left(\frac{M_{\rm fil}}{10^7\,{\rm M_{\odot}}}\right)\left(\frac{v}{100{\rm\,km\,s}^{-1}}\right)\left(\frac{d_{\rm fil}}{\rm kpc}\right)^{-1},
\end{equation} 
where $M_{\rm fil}$ is the total mass of material in the filament.

This total filament mass is uncertain, since the detailed structure and composition of the nebulae are not known. Our optical line flux measurements indicate that the total mass of ionised gas in the filaments of each in these BCGs is less than $\sim10^6$\msol\, (see Table~\ref{tab:totalflux}). However, this probably represents only a fraction of the total mass, since most of the material is likely to exist in cool, dense phases \citep[e.g.,][]{Johnstone:2007p2172}. From \citet{Jaffe:2005p2500}, the approximate mass of warm H$_{2}$ in the filaments of \sersic\, is on the order of \scinot{2}{5}\msol. CO detections imply a mass of $\sim10^{9}$\msol\, of cold molecular gas associated with BCGs of similar luminosity \citep[including \zw;][]{Edge:2001p4022}. 
Assuming that the majority of the material is located in the cool, dense molecular cloud cores in the filaments, the mass in the inner nebula may be $10^{8}-10^{9}$\msol. A significant fraction of this could be located in the bright spiral filaments that dominate the nebulae in \abell\, and \sersic. We adopt a mass of $M_{\rm fil}\approx10^{7}$\msol\, in these filaments, and an average infall velocity of $v=300$\kms. The distances along the filaments are $l_{\rm fil} \sim 7-15$\,kpc for \abell\, and $\sim15-30$\,kpc for \sersic, since the degree of projection is uncertain. Then the minimum mass-inflow rates in the filaments are on the order of several $0.1$\mrate, with infall timescales of $\scinot{2}{7}$ to $10^8$\,yr. 

To consider how these inflows might relate to accretion in the nucleus, we compare the infall rates with estimates of the black hole accretion rates required to fuel the observed radio activity. The 1.4\,GHz radio power of the radio lobes is \scinot{1.4}{24}\,W\,Hz$^{-1}$ in \abell\, \citep{Johnstone:1998p7726}, and \scinot{1.5}{24}\,W\,Hz$^{-1}$ for \sersic\, \citep{Birzan:2008p6520}. Using the radio scaling relation derived by \citet{Birzan:2008p6520} for the radio luminosity at 1.4\,GHz, $L_{\rm1.4\,GHz}$:
\begin{equation}
\log{P_{\rm cav}} = (0.35\pm0.07) \log{L_{\rm 1.4\,GHz}} + (1.85\pm0.10),
\end{equation}
the power required to inflate the lobes $P_{\rm cav}$ in these systems is on the order of $P_{\rm cav} = \scinot{80}{42}$\lum, though there is significant scatter in this relation. The accretion rate $\dot{M}_{\rm acc}$ required for this output is obtained from $\dot{M}_{\rm acc}  = P_{\rm cav} / (\epsilon c^2)$, where $\epsilon$ is the efficiency of the conversion of the gravitational binding energy of the accreting material into jet energy. This is the result of largely unknown physics of AGN jet formation, however a value of $\epsilon\sim0.1$  is commonly adopted \cite[see e.g.,][]{Rafferty:2006p2278}. The required nuclear accretion rate is then approximately $0.01$\mrate\, for both of these galaxies. 
\citet{Rafferty:2006p2278} also estimate accretion rates from energy needed to create the physical volumes of the observed X-ray cavities in these clusters. They obtain black hole accretion rates of $\sim0.003$ and $\sim0.1$\mrate\, for \abell\, and \sersic, respectively. 

That the filament mass infall rates exceed the average black hole accretion rates, particularly given the uncertainty in the parameters, emphasises that these structures could provide sufficient gas flows to power the AGN in these galaxies. However, this is dependent on the efficiency of mass transfer through the inner accretion flows or disks and onto the black hole. 
The results suggest that the filament in \abell, at least, could fuel the central AGN  at the current activity level for $\sim10^7$\,yr. Given the similarity of the rates derived for \sersic, however, mass flow through the filaments could, at some times, limit the fuel for nuclear activity if they are the principal source of material feeding accretion, in this system. 
The timescales for infall are similar to the estimated frequency of AGN outbursts, $10^{7}$ to $10^{8}$ years, which also emphasises that the filaments may be conduits for material inflow during cooling episodes that trigger episodes of activity. In addition, it implies that we might expect to simultaneously observe symptoms of inflow and accretion with those of outflow and heating in some systems, as we do in these galaxies.

In summary, there is evidence in each BCG for an equatorial component of gas rotation in the galaxy nuclei, where the bipolar radio lobes, or outflows, are oriented approximately along the corresponding polar direction. This is consistent with gas accretion in the cores of all three galaxies that may be fuelling nuclear activity and the AGN outflows. It corroborates the model of AGN-powered heating and feedback in the cluster cores, though observations with higher spatial resolution will be required to better constrain the geometry and properties of the flows.

\section{Conclusions} \label{sec:conc}

We have presented optical integral-field spectroscopic observations of the line-emission regions surrounding the BCGs in \abell, \zw, and \sersic. Galaxies such as these are theorized to be participating in a cycle of negative feedback \citep[e.g.,][]{Pizzolato:2005p6412}. This takes the form of energy and mass exchange between the central black hole, host galaxy, and intra-cluster medium, in which the heating that results from the nuclear activity in the galaxy can offset the energy lost to cooling of the thermal gas in the cluster core. The cycle is self-regulating because the AGN output is governed by the amount of mass that cools and is eventually accreted by the black hole.

We have investigated the kinematics and distribution of optical line emission in the extended filamentary nebulae, and compared these properties with the extent and development of the radio emission in the BCG, and with the properties of the ICM. These data provide evidence of the stage of feedback activity that is occurring in these systems, in the framework of the cold feedback scenario.

The data suggest that \abell\, may be in an advanced heating phase of the feedback cycle. The radio lobes are well-formed and appear to have some impact on the surrounding gas, and also to have themselves been influenced  by interactions with the surrounding medium. There is evidence that the expanding radio lobes are evacuating material from the vicinity of the galaxy. Cooling and inflow still appear to be occurring in the nebula, though, with gas possibly on infalling orbital trajectories along the prominent spiral filament in the central nebula. The kinematics of the gas in the galaxy core are consistent with rotation in accretion structures in the nucleus.

\zw\, is likely to be at the opposite end of the feedback cycle, in a phase of strong cooling and accretion, following an episode of AGN activity. A large mass of gas is cooling within a wide region surrounding the galaxy. In this system most of the cooling material appears to be forming a large-scale disk of gas and dust around the central galaxy. There is evidence for rotation and accretion flows on kiloparsec scales in the galaxy core. 

The observations of the optical line emission in \sersic\, show complex structures and kinematics, but do not provide a conclusive picture of the influences on the filaments. As observed in radio and X-ray data, the cluster core has been strongly affected by energy feedback from the central AGN in the galaxy, and likely is experiencing significant heating and only moderate cooling and accretion. There is evidence for gas rotation on scales of several kiloparsecs that may represent accretion flows into the nucleus.

The data for these galaxies appear to be at least qualitatively consistent with a model of cyclic feedback in the cluster cores, so that, as discussed, we can begin to place these systems within the sequence of the feedback cycles that are thought to occur in the cores of galaxy clusters.

\section*{Acknowledgments}
We thank the referee for many helpful suggestions that have significantly improved the paper.
We also acknowledge the Australian Research Council (ARC) for support under Discovery projects DP0342844 (P.M. and C.F.), DP0984657 (M.D.), and DP0664434 (M.D.). M.D. acknowledges the support from the Australian Department of Science and Education (DEST) Systemic Infrastructure Initiative grant and from an Australian Research Council (ARC) Large Equipment Infrastructure Fund (LIEF) grant LE0775546, which together made possible the construction of the WiFeS instrument. 
This research has made use of the NASA/IPAC Extragalactic Database (NED), which is operated by  the Jet Propulsion Laboratory, California Institute of Technology, under contract with the National  Aeronautics and Space Administration, the NASA Astrophysics Data System (ADS), and SAOImage DS9 \citep{Joye:2003p10689}, developed by the Smithsonian Astrophysical Observatory.

\small



%
\begin{deluxetable}{l l c c c c c c } 
\tabletypesize{\scriptsize}
\tablewidth{0pt}
\tablecaption{Properties of the BCG sample\label{tab:sample} }
\tablehead{
\colhead{Cluster ID}	&	\colhead{BCG ID}	&	\colhead{$z$}	&	\colhead{$D$ (Mpc)}	&	\colhead{$L_{\rm H\alpha}$\tablenotemark{a}}	&	\colhead{$L_{\rm rad}$\tablenotemark{b}}	&	\colhead{$L_{\rm X}$\tablenotemark{c}}	&	\colhead{$M_{\rm bulge}$\tablenotemark{d}}	
}
\startdata
\object{A 3581}	&	\object{IC 4374} / PKS 1404-267	&	0.0218	&	93	&	4.7	&	5.4	&	0.27	&	5.7	\\
\object{2A 0335+096}	&	\object{2MASX J03384056+0958119}	&	0.0346	&	147	&	49	&	1.2	&	3.4	&	18	\\
\object{Sersic 159-03}	&	\object{ESO 291-9}	&	0.0564	&	240	&	23	&	21	&	2.2	&	11	\\
A 3526	&	NGC 4696	&	0.0099	&	44	&	1.8	&	8	&	0.28	&	11.6	
\enddata
\tablenotetext{a}{~Integrated total luminosity of H$\alpha$ in $10^{40}$\lum, from this work and \citet{Farage:2010p11714}, assuming the distances in column 4.}
\tablenotetext{b}{~Radio luminosity in $10^{40}$\lum, integrated between 10\,MHz and 10\,GHz, from \citet{Birzan:2008p6520}, except the value for \abell, which is integrated from a power-law fit to the radio flux data available in the NASA Extragalactic Database (NED).}
\tablenotetext{c}{~X-ray luminosity of the cluster inside the cooling radius in $10^{44}$\lum, integrated between 0.001-100\,keV, from \citet{Rafferty:2006p2278}.}
\tablenotetext{d}{~Bulge masses in $10^{11}$\msol, from \citet{Rafferty:2006p2278}, based on the $R$-band magnitudes.}
\end{deluxetable}
\begin{deluxetable}{l c  l cc}
\tabletypesize{\scriptsize}
\tablewidth{0pt}
\tablecaption{Summary of WiFeS observations\label{tab:obs}}
\tablehead{
\colhead{Host cluster}	&  \colhead{Scale} &	\colhead{Dates observed (2009)}	& \colhead{ Total exp. time }	&\colhead{PA} \\
\colhead{} &  \colhead{(kpc\,arcsec$^{-1}$)} & \colhead{}  &  \colhead{(hr)} &  \colhead{(deg)} 
}
\startdata
Abell 3581	&	0.455	&  May 1-3;  23-24  			&  11.5	&  180	\\
2A 0335+096 	&	0.675	&  Oct 15-21; Nov 13-15  		&  11.0	&  330	\\
Sersic 159-03	&	1.074	&  Jul 20, 23-24; Oct 16, 21	&  11.5	&  188	
\enddata
\end{deluxetable}
\begin{deluxetable}{ lcc ccc ccc } 
\tabletypesize{\scriptsize}
\tablewidth{0pt}
\tablecaption{Total \ha\, and [\ion{N}{2}] fluxes and luminosities\label{tab:totalflux}}
\tablehead{
\colhead{Cluster ID}	 &  \colhead{$A_{V, {\rm gal}}$\tablenotemark{a}}  &  \colhead{$A_{V, {\rm avg}}$\tablenotemark{b}}  &  \colhead{$F({\rm H\alpha})$}  &  \colhead{$L({\rm H}\alpha)$}  & \colhead{$M$(\ion{H}{2})}  &  \colhead{$F$([\ion{N}{2}])}  &  \colhead{$L$([\ion{N}{2}])}   & \colhead{[\ion{N}{2}]/H$\alpha$}
}
\startdata
A 3581	&	0.20	&	0.4	&	$4.5\pm0.5$	&	$4.7\pm0.5$	& $<\scinot{5.5}{5}$	&	$6.8\pm0.7$	&	$7.0\pm0.8$	&  $1.5\pm0.2$  \\
2A 0335+096	&	1.4	&	1.5	&	$19\pm2$\phantom{ \,}	&	$49\pm6$\phantom{\, }	& $<\scinot{5.8}{6}$	&	$18\pm2$\phantom{\, }	&	$47\pm6$\phantom{\, } & $1.0\pm0.2$	\\
Sersic 159-03	&	0.04	&	0.5	&	$3.3\pm0.4$	&	$23\pm3$\phantom{\, } &	$<\scinot{5.7}{6}$	&	$2.9\pm0.3$	&	$20\pm2$\phantom{\, }  & $0.9\pm0.1$
\enddata
\tablenotetext{~}{~Line fluxes are in units of $10^{-14}$\flux, and luminosities in $10^{40}$\lum. Column 5 shows the estimated upper limit on the mass of ionized hydrogen in \msol, for an electron density  $n_{\rm e} < 200$\cc. }
\tablenotetext{a}{~Foreground Galactic extinction (in magnitudes) from \citet{Schlegel:1998p10426}. }
\tablenotetext{b}{~Extinction (in magnitudes) assumed in pixels where it can not be estimated from the H$\alpha$ and H$\beta$ fluxes and the assumed extinction law.}
\end{deluxetable}
\begin{deluxetable}{ l cc cc cc } 
\tabletypesize{\scriptsize}
\tablewidth{0pt}
\tablecolumns{7}
\tablecaption{Extinction-corrected line fluxes in selected 5\arcsec-diameter apertures \label{tab:linefluxes}}
\tablehead{
	& \multicolumn{2}{c}{\abell}			& \multicolumn{2}{c}{\zw} &	\multicolumn{2}{c}{\sersic } \\
	& \colhead{Nucleus (0,0)}	& \colhead{Filament (-4,-4)\tablenotemark{a}} & \colhead{Nucleus (0,0)}	& \colhead{Filament (1,17)\tablenotemark{a}}	& \colhead{Nucleus (0,0)} & \colhead{Filament (-1,4)\tablenotemark{a}}
}
\startdata
$F({\rm H}\beta)$\tablenotemark{b}	&	$3.8\pm0.4$	&	$1.5\pm0.2$	&	$8\pm1$	&	$4.1\pm0.5$	&	$2.3\pm0.3$	&	$2.2\pm0.3$	\\
\cutinhead{Emission-line strength relative to $F({\rm H}\beta)$} 
{[O II]} $\lambda$3726+3729	&	$4.9\pm0.2$	&	$6.6\pm0.2$	&	$6.1\pm0.2$	&	$6.0\pm0.4$	&	$6.3\pm0.2$	&	$5.0\pm0.1$	\\
{[O III]} $\lambda$5007	&	$2.16\pm0.07$	&	$0.77\pm0.03$	&	$<0.63$	&	$<0.58$	&	$<0.54$	&	$<0.37$	\\
{[N I]} $\lambda$5198+5200	&	$<0.33$	&	$<0.40$	&	$<0.29$	&	$<0.54$	&	$<0.31$	&	$<0.37$	\\
{[O I]} $\lambda$6300	&	$0.66\pm0.04$	&	$0.57\pm0.03$	&	$0.75\pm0.02$	&	$0.61\pm0.03$	&	$0.64\pm0.03$	&	$0.60\pm0.02$	\\
H$\alpha$	&	$4.2\pm0.1$	&	$3.9\pm0.1$	&	$3.10\pm0.07$	&	$2.9\pm0.1$	&	$3.1\pm0.1$	&	$3.10\pm0.08$	\\
{[N II]} $\lambda$6583	&	$6.9\pm0.2$	&	$5.7\pm0.2$	&	$3.49\pm0.07$	&	$2.8\pm0.1$	&	$3.4\pm0.1$	&	$2.70\pm0.07$	\\
{[S II]} $\lambda$6717	&	$2.61\pm0.08$	&	$1.74\pm0.05$	&	$1.61\pm0.04$	&	$0.85\pm0.04$	&	$1.47\pm0.05$	&	$0.89\pm0.03$	\\
{[S II]} $\lambda$6731	&	$1.93\pm0.06$	&	$1.19\pm0.04$	&	$0.98\pm0.02$	&	$0.58\pm0.03$	&	$0.88\pm0.03$	&	$0.58\pm0.02$	
\enddata	
\tablenotetext{a}{~Coordinates of the aperture centers are given as offsets from the galaxy nucleus in arcseconds.}
\tablenotetext{b}{~H$\beta$ line flux in units of $10^{-15}$\flux.}
\end{deluxetable}
%


%

\end{document}